\documentstyle[12pt]{report} 
\def\rbuildrel#1\over#2{\mathrel{\mathop{#2}\limits_{#1}}}

\catcode`@=11
\def\underline#1{\relax\ifmmode\@@underline#1\else
        $\@@underline{\hbox{#1}}$\relax\fi}

\def\titlepage{\pagestyle{empty}\c@page=0
      \def\thefootnote{\fnsymbol{footnote}} }
\def\endtitlepage{\pagestyle{plain}\c@page=1
      \def\thefootnote{\arabic{footnote}} \c@footnote\z@ }
\catcode`@=12

\newskip\humongous \humongous=0pt plus 1000pt minus 1000pt

\newif\ifdtup

\def\*{\hskip .06 cm}

\def\thebibliography#1{\section*{\ \markboth
 {REFERENCES}{REFERENCES}}\list
 {{\arabic{enumi}}.}
 {\settowidth\labelwidth{{#1}.}\leftmargin\labelwidth
 \advance\leftmargin\labelsep
 \usecounter{enumi}}
 \def\newblock{\hskip .11em plus .33em minus -.07em}
 \sloppy
 \sfcode`\.=1000\relax}

\def\thebibliographyp#1{\section*{\ \markboth
 {Chan \& Dill, $\quad$ Polymer Principles in Protein Structure
 and Stability}{Chan \& Dill, $\quad$ Polymer Principles in Protein Structure
 and Stability}}\list
 {\arabic{enumi}.}{\settowidth\labelwidth{#1.}\leftmargin\labelwidth
 \advance\leftmargin\labelsep
 \usecounter{enumi}}
 \def\newblock{\hskip .11em plus .33em minus -.07em}
 \sloppy
 \sfcode`\.=1000\relax}

\def\sqr#1#2{{\vcenter{\vbox{\hrule height.#2pt
\hbox{\vrule width.#2pt height#1pt \kern#1pt
\vrule width.#2pt}
\hrule height.#2pt}}}}

\begin{document}
%
\noindent
$\null$
\hfill November 29, 2002\\

\vskip 0.6in 

\begin{center}
{\Large\bf Solvation Effects and Driving Forces for Protein}\\

\vskip 0.3cm

{\Large\bf Thermodynamic And Kinetic Cooperativity: How}\\ 

\vskip 0.3cm

{\Large\bf Adequate Is Native-Centric Topological Modeling?}\\

\vskip .5in 
{\bf H\"useyin K{\footnotesize{\bf{AYA}}}}
and
{\bf Hue Sun C{\footnotesize{\bf{HAN}}}}$^\dagger$\\
$\null$

Protein Engineering Network of Centres of Excellence (PENCE),\\
Department of Biochemistry, and
Department of Medical Genetics \& Microbiology,
Faculty of Medicine, University of Toronto,
Toronto, Ontario M5S 1A8, Canada\\

%

%

\end{center}

\vskip 1cm

\noindent
{\bf Running title:} Continuum G\=o Model Chevron Plots\\

\vskip 1cm

chevron plot / desolvation barrier /
continuum G\=o models / heat capacity

$\null$\\ 
%


\noindent 
$^\dagger$ Corresponding author.\\ 
E-mail address of Hue Sun C{\footnotesize{HAN}}: 
chan@arrhenius.med.toronto.edu\\
Tel: (416)978-2697; Fax: (416)978-8548\\
Mailing address: Department of Biochemistry, University of Toronto, 
Medical Sciences Building -- 5th Fl., 1 King's College Circle,
Toronto, Ontario M5S 1A8, Canada.

\vfill\eject 
%

\def\thefootnote{\fnsymbol{footnote}}

\noindent 
{\large\bf Summary}\\

\vskip .2 in

\noindent 
What energetic and solvation effects underlie the remarkable two-state 
thermodynamics and folding/unfolding kinetics of small single-domain 
proteins? To address this question, we investigate the folding and 
unfolding of a hierarchy of continuum Langevin dynamics models of 
chymotrypsin inhibitor 2. We find that residue-based additive G\=o-like 
contact energies, although native-centric, are by themselves insufficient for 
proteinlike calorimetric two-state cooperativity. Further native biases 
by local conformational preferences are necessary for proteinlike 
thermodynamics. Kinetically, however, even models with both 
contact and local native-centric energies do not produce simple two-state 
chevron plots. Thus a model protein's thermodynamic cooperativity 
is not sufficient for simple two-state kinetics. The models tested appear to 
have increasing internal friction with increasing native stability, 
leading to chevron rollovers that typify kinetics that are commonly
referred to as non-two-state. The free energy profiles of these models are 
found to be sensitive to the choice of native contacts and the presumed
spatial ranges of the contact interactions. Motivated by 
explicit-water considerations, we explore recent treatments 
of solvent granularity that incorporate desolvation free energy barriers 
into effective implicit-solvent intraprotein interactions.
This additional feature reduces both folding and unfolding rates 
vis-\`a-vis that of the corresponding models without desolvation barriers, 
but the kinetics remain non-two-state. Taken together, our 
observations suggest that interaction mechanisms more intricate than 
simple G\=o-like constructs and pairwise additive solvation-like contributions 
are needed to rationalize some of the most basic generic protein 
properties. Therefore, as experimental constraints on protein chain 
models, requiring a consistent account of proteinlike thermodynamic 
and kinetic cooperativity can be more stringent and productive for
some applications than simply requiring a model heteropolymer to fold 
to a target structure.

\vfill\eject 

\centerline{\bf INTRODUCTION}

$\null$

A fundamental unresolved question in molecular biology is how 
solvent-mediated interactions conspire to produce the highly 
specific structures and dynamics of proteins. 
Recent experiments on highly cooperative ``two-state'' folding/unfolding 
kinetics of small single-domain proteins$^{1-4}$ have, however, revealed
an intriguing phenomenological simplicity. Most notably, the folding 
rates of these proteins are found to be well correlated with a simple 
contact order parameter deducible entirely from the native contact 
pattern, often referred to as a protein's ``topology.''$^{1,5-7}$
\\

From a reductionist viewpoint, protein behavior is ultimately determined 
by the large collection of atoms of the protein and those that constitute 
the solvent. Indeed, computational studies of protein folding by all-atom 
simulations have led to many useful insights.$^{8-10}$ As well, recent 
developments in Monte Carlo conformational sampling techniques are 
promising (see, e.g., ref.~10). However, currently even the most extensive 
all-atom explicit-solvent simulations of proteins$^{8,9}$ do not appear to 
provide sufficient conformational coverage to tackle many equilibrium and 
long-timescale kinetic properties under ambient conditions. Thus, a 
long-standing$^{11-13}$ complementary approach is to adopt simplified 
lattice$^{14-19}$ or continuum$^{19-21}$ representations of polypeptide 
geometries and interactions, trading high structural resolution for the 
extensive conformational sampling that is often necessary for conceptual 
advances.$^{15-21}$ 
\\

Considerable progress has been made using simplified representations. But
still it has been exceedingly difficult to fold a heteropolymer chain to the 
native structure of a real protein by a potential function determined solely 
by the chain's one-dimensional amino acid sequence. In this context, the 
recent discovery of certain predictive powers of native topology$^{5}$ has 
inspired intense interests in native-centric modeling.$^{22-46}$ These 
models are often called G\=o or G\=o-like since they make no explicit 
reference to a protein's sequence. Instead, teleological$^{47,48}$ interaction 
schemes are postulated to bias conformations toward a given native structure, 
in a manner similar to the original lattice constructs of G\=o and 
coworkers.$^{12,49}$ 
\\

Remarkably, despite their apparent simplicity and artificiality, tremendous 
advance has been made by recent innovations in native-centric
modeling. This investigative approach has proven to be a very effective 
tool to gain understanding in the face of difficulties encountered 
by more reductionist approaches.$^{50}$ Through the questions they posed, 
critical physical insights have been gained into the 
thermodynamics,$^{27,28,30,32,37-41,45}$ 
folding kinetics,$^{30-35,38-46}$ folding rates,$^{22-25,35,43}$ 
and transition states and folding 
interme\-diates$^{26,31,32,36,40,42,43,46}$ of proteins.
For example, physical rationalizations of the observed relationship 
between contact order and folding rate have been provided by Ising-type 
native-centric models without an explicit chain representation$^{23-25}$
as well as explicit G\=o-like chain models.$^{35}$
In separate efforts, using a protein's database native structure as
starting point, Gaussian elastic network$^{51,52}$ and graph
theoretic$^{53}$ models have been notably successful in deciphering
the flexibility and vibrational modes of real proteins without 
explicitly considering the myriad intraprotein interactions involved.
These models of near-native dynamics do not originally tackle 
protein folding/unfolding kinetics. But in a recent generalization
of the Gaussian elastic network model that took into account chain 
connectivity, a significant correlation between experimental folding 
rates and the relaxation rates of the slowest vibrational modes was 
discovered,$^{44}$ suggesting an intimate connection between near-native
vibrations and folding/unfolding kinetics. 
\\

Some of the successes of native-centric approaches have 
been attributed to the stipu\-lation$^{28,32}$ 
that G\=o-like potentials are proteinlike to a certain degree because 
they serve to eliminate ``to first order''$^{28}$ or to minimize$^{32}$ the 
``energetic frustration'' that is presumed to be minimized in real 
proteins.$^{54-56}$ According to this view, native-centric models are 
thus left free to account for ``topological frustration''$^{28,32}$ alone ---
i.e., to capture the physics arising solely from chain connectivity,
excluded volume, and the favorability of the native fold. 
\\

While we cherish the successes of the native-centric paradigm,
it is also important to not lose sight of its limitations. 
In short, native-centric modeling entails: (i) Admitting our ignorance of 
the basic physics of protein folding, at least for the time being. 
(ii) Recognizing that a protein sequence's known native structure 
may contain significant information about its actual energetics.
(iii) Assuming that a G\=o-like potential inferred from the known native 
structure is in some sense an approximate description of the underlying 
physical energetics. (iv) Working out the logical consequences 
of these assumptions to gain insight into various aspects of the folding 
process. In this perspective, native-centric modeling should be taken as a 
tentative means to capture collective atomic/molecular effects that we don't 
yet understand. As such, its application may be physically more meaningful 
at a coarse-grained level (perhaps as a ``renormalized'' description). 
Although all-atom G\=o-like models$^{36,46}$ are obviously superior in 
accounting for the important effects of sidechain packing (see, e.g.,
insightful discussion in ref.~46), physically it is even harder
to justify why the interaction between a pair of atoms in general would 
depend on whether they are in close spatial proximity in a particular 
protein's native structure. Historically speaking, the renewed 
popularity of G\=o-like models in protein folding studies since the late 
1990s may arguably represent a partial backtracking --- albeit a very 
productive and well justified one --- in modeling philosophy. This is so 
because the desire to supersede these earlier ad hoc interaction schemes 
appeared to be an impetus for the emergence since the late 1980s of 
simplified lattice protein folding models with general sequence-dependent 
potentials.$^{57-62}$ 
\\

Here we endeavor to better delineate the utility and limitation of 
several common native-centric approaches to protein folding.
In identifying their strengths as well as weaknesses, our goal is
to pave the way for improved native-centric modeling and better
reductionist approaches. It goes without saying that
G\=o-like models are intrinsically incomplete, because (i) possible 
nonnative interactions are in large measure neglected,$^{63,64}$ and (ii) 
interaction (energetic) heterogeneities can be present in proteins with 
essentially identical topologies.$^{65}$ Practically speaking, even if 
a general usefulness of native-centric modeling is presumed, robustness 
of the predictions has to be ascertained. Many native-centric interaction 
schemes can be postulated; not all of them have the same predictions. Some 
discrepancies are puzzling.  For example, the combination of a G\=o-like 
potential with an explicit chain representation should, in principle, be a 
more adequate model of topological frustration$^{28,32}$ than models without 
an explicit chain representation. Yet so far native-centric constructs 
with geometrically less realistic (non-explicit) chain 
representations$^{23-25,43}$ appear to be more successful in reproducing 
experimental folding rates than direct folding kinetics simulations of 
G\=o-like models with explicit chain representations.$^{35}$ 
\\

Specifically, the present work addresses two basic questions of robustness 
in native-centric modeling: (i) How much do the model predictions depend 
on the choice of native contacts for a given protein? (ii) To what extent 
would these predictions be modified when the effects of the protein's 
aqueous solvent are taken into account with more sophistication?$^{40,66}$ 
Pursuing a line of inquiry we have recently developed in the context
of lattice models,$^{18,47,48,67,68}$ we focus here on whether continuum 
coarse-grained G\=o-like energetics with an explicit chain representation 
can reproduce certain generic thermodynamic 
and kinetic cooperativities that have been experimentally observed across many 
real proteins. These statistical mechanics tests are stringent. For 
instance, the mere existence of a qualitatively 
sharp folding transition in a chain model does not necessarily imply that 
its underlying thermodynamics is proteinlike.$^{18,47,48,67}$ 
Homopolymers can have very sharp coil-globule transitions that are
not calorimetrically two-state.$^{69}$ 
Comparisons between simulated and experimental chevron 
plots$^{70}$ show that even for chain models that satisfy the experimental 
thermodynamic two-state criteria, it is nontrivial$^{68}$ to reproduce 
the highly cooperative nonglassy two-state folding kinetics$^{71}$ of many 
small single-domain proteins. Therefore, applying these tests would, in 
due course, facilitate the improvement of existing models, suggest yet 
unexplored avenues of native-centric topological modeling, and ultimately 
help decipher the energetics of real proteins.
\\


$\null$

\centerline{\bf COMPARING DIFFERENT NATIVE CONTACT SETS FOR CI2}

$\null$

We consider the 64-residue truncated form 
of chymotrypsin inhibitor 2 (CI2)$^{72}$ using coarse-grained C$_\alpha$ 
representations with sidechain interactions accounted for by contacts 
between pairs of C$_\alpha$ positions separated by at least three 
C$_\alpha$s along the chain sequence (contact order $\ge 4$). CI2 is a widely 
studied small single-domain protein with no disulfide bond. It folds and 
unfolds as an apparent simple two-state system. CI2 is an ideal test case 
because a large body of experimental, all-atom molecular dynamics, and 
native-centric modeling data is available (see, e.g., refs.~32, 36, 
73, 74). To investigate how coarse-grained native-centric model 
predictions may be sensitive to the definition of native contacts, here 
we examine two native contact sets, which we refer to as NCS1 and NCS2.
\\

NCS1 is determined by the distance criterion of Shea et al:$^{28}$ 
Two amino acid residues $i$ and $j$ of a given protein are in contact 
if, in its native structure from the Protein Data Bank (PDB), either
their C$_{\alpha}$ atoms are less than 8\AA~ apart, or any two
heavy atoms one from each of their two sidechains are less than 4\AA~ 
apart, or both. Using this definition, there are 137 NCS1 contacts. 
NCS2 is borrowed from Clementi et al.'s native contact map for CI2 
(Figure~2 of ref.~32). NCS2 has 142 contacts.
It was based upon the CSU software$^{75}$ which takes into account more 
detailed structural information\footnote{
For CI2, the current version of the CSU software available from the 
Internet also produces a set of 142 native contacts, all except 8 of which
are identical to the contacts in NCS2. For the computational tests
we have conducted (detailed data not shown), differences in results 
for this particular CSU native contact set and that for NCS2 are negligible.
}
such as contact surface area and solvent accessibility.
There are considerable variations in native C$_\alpha$--C$_\alpha$ distances 
among contacts in both NCS1 and NCS2. The minimum native contact distance 
is 4.325 \AA~ for both sets, but the maximum are 
12.255 \AA~ and 15.558 \AA~ for NCS1 and NCS2 respectively. The
average native distance of NCS1 (6.528 \AA) is smaller than that of
NCS2 (7.288 \AA). However, the average sequence separations of NCS1 
($23.1$) and NCS2 ($22.6$) are almost identical. 
\\

Figure~1 compares the two native contact sets. They have 108 contacts
in common (blue lines in Figure~1b). Among the native contacts that are 
not common to both sets, those belonging to NCS1 but not NCS2 (green lines 
in Figure~1c) tend to be between two ends of the chain or involve the 
$\beta 1$ strand (residues 27--34).$^{73}$ In contrast, contacts 
belonging to NCS2 but not NCS1 (red lines in Figure~1d) appear to be
more uniformly distributed, involving more the $\alpha$-helix
(residues 13--23) and the region spanning residues 35--44. Specific
examples of these differences are provided in Figure~2, showing 
that NCS1 identifies an hydrophobic-polar 
(alanine-arginine) contact but not an hydrophobic-hydrophobic 
(valine-phenylalanine) contact.
\\


$\null$


\centerline{\bf MODELS AND METHODS}

$\null$

\noindent
{\large\it Coarse-grained Potentials Without Solvation/Desolvation Barriers}

The basic construct of our native-centric potentials follows that
of Clementi et al.$^{32}$ For a given model protein conformation 
specified by the positions of all its C$_\alpha$ atoms, the total 
G\=o-like potential energy 
\begin{eqnarray}
V_{{\rm total}}
&=&V_{{\rm stretching}}+
V_{{\rm bending}}+
V_{{\rm torsion}}+
V_{{\rm non-bonded}} \nonumber\\
  &=&\sum_{{\rm bonds}}^{N-1} K_r (r-r_0)^2+
\sum_{{\rm angles}}^{N-2} K_{\theta} (\theta-\theta_0)^2 \nonumber \\
& &+\sum_{{\rm dihedrals}}^{N-3}\Bigl\{ K_{\phi}^{(1) } 
[1-\cos (\phi-\phi_0)] +
K_{\phi}^{(3)} [1-\cos 3(\phi-\phi_0)] \Bigr \} \nonumber \\
& &+\sum_{i<j-3}^{{\rm native}}
\epsilon\bigg[ 5\bigg(\frac{r^\prime_{ij}}{r_{ij}} \bigg)^{12}
-6\bigg(\frac{r^\prime_{ij}}{r_{ij}} \bigg)^{10}
\bigg]
+\sum_{i<j-3}^{{\rm non-native}} \epsilon \bigg(\frac{r_{{\rm rep}}}{r_{ij}}
\bigg)^{12} \; ,
\end{eqnarray}
where $N$ is the total number of particles.
This functional form has also been used by Koga and 
Takada.$^{35}$ Here the first three summations are for local 
interactions, where $r$, $\theta$, and $\phi$ are, respectively, 
the C$_\alpha$--C$_\alpha$ virtual bond length between successive 
residues along the chain sequence, C$_\alpha$--C$_\alpha$ virtual 
bond angles, and C$_\alpha$--C$_\alpha$ virtual torsion angles;
$r_0$, $\theta_0$, and $\phi_0$ are the 
corresponding native values in the PDB structure.
These terms account for chain connectivity and presumed local 
conformational preferences for the native fold. 
The last two summations are for nonlocal interactions;
$r_{ij}$ is the spatial distance between two C$_\alpha$s
that have at least three residues between them along the chain 
sequence. In the summation over native contacts (as defined above
for either NCS1 or NCS2), a $10$--$12$ Lennard Jones (LJ) form
is used, where $r^\prime_{ij}$ is the C$_\alpha$--C$_\alpha$ 
distance between the contacting residue $i$ and residue $j$ in the 
PDB structure. In the summation over non-native contacts,
$r_{{\rm rep}}$ parametrizes the excluded volume repulsion between
residue pairs that do not belong to the given native contact set.
As in refs.~32 and 35,
we use $r_{{\rm rep}}=4$ \AA~
(whereas ref.~28 uses $r_{{\rm rep}}=7.8$ \AA).
The ratios between interaction parameters are
$K_r=100\epsilon$, $K_{\theta}=20\epsilon$, $K_{\phi}^{(1)}=\epsilon$, and 
$K_{\phi}^{(3)}=0.5\epsilon$, as in ref.~32. 
The interaction strength is thus controlled by a single 
parameter $\epsilon$. We refer to the potential just described 
as the ``without-solvation'' model because it does not have a 
solvation/desolvation barrier (see below), although the terms 
in equation~1 may be interpreted as part of an implicit-solvent
scheme that takes into account other aspects of solvent-mediated 
interactions.
\\

Equation~1 assumes that native-centric favorable interactions have
relatively long spatial ranges. In alternate square-well G\=o
models,$^{41}$ however, favorable contact interactions have sharp cutoffs.
Moreover, in many lattice models, contact interactions may be viewed as 
having infinitesimal spatial ranges. Thus, to investigate how the presumed 
spatial ranges of contact interactions may affect model predictions, we study 
a variation of the above model that restricts each of the pairwise 
$10$--$12$ LJ native contact terms in equation~1 to $r_{ij}\le 1.2 
r^\prime_{ij}$ and sets the interaction to zero for $r_{ij}> 1.2 
r^\prime_{ij}$, but all other aspects of the model stay the same. We call 
this the ``without-solvation-SSR'' (short spatial range) model.
\\


\noindent
{\large\it An Approximate Account of Solvation/Desolvation Barriers}

We consider also coarse-grained ``with-solvation'' models designed 
to semi-quantitatively account for the
solvation/desolvation free energy barriers encountered by a 
protein's constituent groups as they cluster together in aqueous 
solvents (Figure~3). We refer to these barriers simply as 
``desolvation barriers'' below. Desolvation barriers are a robust 
consequence of granularity or the particulate nature of 
the solvent.$^{76}$ They have
long been predicted by theory$^{77}$ and atomic simulations.$^{78,79}$ 
However, aside from an earlier study that used a square-well/square-shoulder 
form of desolvation barriers,$^{66}$ until very recently$^{40,80,81}$
this salient physical feature was not taken into account in continuum 
coarse-grained protein models. While explicit-solvent molecular dynamics 
account for solvation effects directly, these simulations do not yet 
provide a definitive answer as to whether they can or cannot reproduce 
the experimentally observed thermodynamic and kinetic cooperativities 
in protein folding. Therefore,
complementary ``implicit-solvent''$^{82}$ treatments$^{40,66,79,80}$ like
the present approach are
needed. Indeed, the experimentally based cooperative tests 
conducted here should also be applied to all-atom models once their 
computational efficiency has improved to make it possible.
\\

The scope of the present work is limited. In particular, the study of 
structural details --- such as connections to the powerful experimental
$\Phi$-value analysis of transition-state structures,$^{4,72,74}$ is deferred 
to future applications of our investigative framework. 
We first tackle a little-explored but fundamental 
question: How deeply are 
protein folding thermodynamic and kinetic cooperativities affected by the 
introduction of generic desolvation barriers?
To this end, we employ the general implicit-solvent functional form introduced 
recently by Cheung et al.$^{40}$ (Figure~3). The repulsive part of this 
potential (for $r<r^\prime$) is similar, though not identical, to the repulsive 
part of the $10$--$12$ LJ term in the without-solvation model 
above (equation~1).
The key difference is that now a free energy barrier is present at the 
midpoint $(r^\prime+r^{\prime\prime})/2$ between the contact
($r^\prime$) and water-separated ($r^{\prime\prime}$) free energy minima 
of a given pairwise interaction; $r^{\prime\prime}-r^\prime=3.0$ \AA~ is
the approximate diameter of a single water molecule.
Shown in Figure~3b(i) is a potential with relative magnitudes 
of the barrier and minima similar to that in ref.~40.
This form has a relatively high desolvation barrier.\footnote{
In order not to have a negative $\epsilon^{\prime\prime}/\epsilon$ ratio,
it appears that the relation 
$(\epsilon^{\prime\prime}-\epsilon^\prime)/(\epsilon^\prime-\epsilon)$ $=1.33$ 
in the legend for Figure~1 in Cheung et al.$^{40}$
should read
$(\epsilon^{\prime\prime}+\epsilon^\prime)/(\epsilon^\prime-\epsilon)$ $=1.33$. 
}
The $U(r)$ function in the present study has a lower 
barrier (Figure~3a). As our goal is only to elucidate the 
generic implications of having a significant desolvation barrier,
provided that the barrier is not negligible, a lower barrier is
advantageous because it allows for faster 
kinetics and thus broader conformational sampling. Not the
least, our choice is not inconsistent with recent explicit-water atomic 
simulations that predicted a lower pairwise desolvation barrier$^{83}$ 
[Figure~3b(ii)]. Now, for the with-solvation model, we simply replace 
the pairwise $10$--$12$ LJ terms of the second last summation over native 
contacts in the $V_{\rm total}$ equation~1 above with $U(r)$s (Figure~3a) 
for the corresponding native pairs. Other terms in equation~1 remain the 
same. We call the resulting potential function $V_{\rm total}^{(S)}$.  
Again, the interaction strength of a given model is controlled by 
one single parameter $\epsilon$. In principle, terms in both the
without-solvation and with-solvation potentials 
representing solvent-mediated interactions can depend on
temperature.$^{84-86}$ To simplify the formulation,
however, and especially since most of the results in this report 
entail comparing kinetic trajectories under a constant given temperature,
here $V_{\rm total}$ and $U(r)$ are taken to be
temperature independent, as in refs.~32 and 40.
\\

\noindent
{\large\it Langevin Dynamics}

Folding and unfolding kinetics are simulated by 
Langevin dynamics,\footnote
{Alternately, Newtonian dynamics in conjunction with the Berendsen et al. 
algorithm$^{87}$ for coupling to a heat bath was used by
several previous investigations$^{32,35}$ of 
similar G\=o-like coarse-grained protein models.}
using a formulation similar to Thirumalai and coworkers'.$^{88,89}$
For each of the $3N$ degrees of freedom of the model protein 
($x$, $y$ or $z$ coordinates of the C$_\alpha$s), the equation 
of motion is:
\begin{equation}
m \dot{v}(t)=F_{{\rm conf}}(t)-m\gamma v(t)+ \eta(t) \; ,
\end{equation}
where $m$, $v$, $\dot{v}$, $F_{{\rm conf}}$, $\gamma$ and $\eta$ are,
respectively, mass, velocity, acceleration, conformational force,
friction (viscosity) constant and random force. The conformational
force is equal to the negative gradient of the total potential energy 
of the given model ($V_{\rm total}$ or $V_{\rm total}^{(S)}$). For
the without-solvation-SSR models, conformational force from
the pairwise $10$--$12$ LJ native contact term in $V_{\rm total}$
between residues $i$ and $j$ is applied only 
if $r_{ij}\le 1.2 r^\prime_{ij}$.  The random force has the 
autocorrelation function
\begin{equation}
\langle \eta(t)\eta(t^\prime) \rangle =
2m\gamma\, k_{\rm B}T\, \delta (t-t^\prime) \; ,
\end{equation}
where $k_{\rm B}T$ is Boltzmann constant times absolute temperature. 
Every C$_\alpha$ is subject to a random force at each integration time step.
The components of the random force are independently 
generated by setting $\eta_i=(2m\gamma k_{\rm B}T/\delta t)^{1/2}\xi_i$. 
Here $i$ denotes the uncorrelated random force components in the
$x$, $y$ or $z$ directions, $\xi_i$ is a random value taken from a Gaussian 
distribution with zero mean and unit variance (obtained from a random
number generator by standard techniques$^{90}$), and $\delta t$ 
is the integration time step. At the commencement of a simulation at
temperature $T$, the initial velocities are assigned random values 
by setting $v_i=(k_BT/m)^{1/2}\xi_i$.

We use the velocity-verlet algorithm$^{88-91}$
(equations~12 and 13 in ref.~89) to integrate equation~2. 
Independent of simulation conditions such as variations in $\epsilon$ and 
$T$, the time scale of the model systems here is always controlled by 
the quantity $\tau=\sqrt{ma^2/\epsilon_0}$, with the length scale $a=4$ \AA~ 
and a reference energy scale
$\epsilon_0=1$. We further set $\gamma=0.05\tau^{-1}$ and use a molecular
dynamics time step $\delta t=0.005 \tau$ in the numerical integration. 
Conformational sampling is performed by averaging over snapshots taken at 
every $400$ time steps. Simulation times in this study
are presented in units of $\delta t$. The energy parameter $\epsilon$
and temperature $T$ are given respectively in units of $\epsilon_0$ and 
$\epsilon_0/k_{\rm B}$, and length is measured in units of \AA. To simplify 
notation, other units are chosen such that $m=1$ and $k_{\rm B}=1$ 
in the present simulations, as in Veitshans et al.$^{89}$ 
An approximate correspondence between model time 
and real protein kinetic time scales can be found in ref.~89.
\\


$\null$

\centerline{\bf THERMODYNAMIC COOPERATIVITY}

$\null$

\noindent
{\large\it Free Energy Profiles in Different Native-Centric Schemes}

Using the progress variable $Q$ (native contact fraction), 
Figure~4 shows that conformational 
distribution is significantly sensitive to the choice of native 
contact set and the presumed spatial ranges of native contact interactions. 
Consistent with the expectation for a two-state protein
such as CI2 and a previous without-solvation study,$^{32}$ the free 
energy profiles for NCS2 (solid curves) exhibit
a single peak at intermediate $Q$ separating the native (high $Q$) and 
denatured (low $Q$) minima. In contrast, the NCS1 free energy 
profile has a plateau-like transition region in the without-solvation 
formulation (Figure~4a, dashed curve). More remarkably, for the 
without-solvation-SSR and with-solvation models (Figure~4b, c), the NCS1 
profiles develop a shallow minimum flanked by two peaks in the intermediate 
$Q$ region (dashed curves), similar to certain postulated free energy 
profiles discussed previously, for example, by Fersht$^{92}$ and 
Chu and Bai,$^{93}$ in the context of folding kinetics that apparently 
involves intermediates. Also notable is the progressive movement of 
the native minimum position from $Q\approx 0.9$ for the without-solvation 
models toward $Q=1$ for the with-solvation models.
The incorporation of desolvation barriers 
dramatically raises the overall folding/unfolding free energy barrier for 
NCS2, but only has a relatively subdued effect for NCS1 (c.f. Figure~4b, c), 
suggesting that there is an intricate interplay between desolvation
barrier effects and other aspects of solvent-mediated 
interactions in protein folding.
\\

\noindent
{\large\it Calorimetric Cooperativity: Local Conformational Preferences 
are Crucial}

Figure~5 assesses the calorimetric cooperativity$^{47,48,67,68}$ of seven
different native-centric models of CI2 by comparing their simulated 
van't Hoff over calorimetric enthalpy ratios 
$\Delta H_{\rm vH}/\Delta H_{\rm cal}$ to the experimental two-state
requirement that $\Delta H_{\rm vH}/\Delta H_{\rm cal}\approx 1$.
Model intraprotein interactions are taken to be temperature 
independent in this evaluation. Since vibrations along the bonds (equation~1) 
contribute to heat capacity in these models outside the 
folding/unfolding transition region, and there is experimental
evidence for heat capacity contributions from bond vector motion in real 
proteins,$^{94}$ the simulated $\Delta H_{\rm vH}/\Delta H_{\rm cal}$ ratio
without baseline subtractions does not correspond physically to the 
experimental $\Delta H_{\rm vH}/\Delta H_{\rm cal}$ ratio obtained 
by empirical baseline subtractions.$^{47,48,67,68}$ Thus, only 
the baseline-subtracted $\Delta H_{\rm vH}/\Delta H_{\rm cal}$ ratio
$\kappa_2^{\rm (s)}$ from the models are judged against the 
experimental calorimetric two-state criterion.$^{47,48,68}$ 
Figure~5 shows that 
$\Delta H_{\rm vH}/\Delta H_{\rm cal}$ $=\kappa_2^{\rm (s)}$ $\approx 1$ 
is satisfied by all six models described in the last section.
Apparently,
similar G\=o-like models in refs.~32 and 40
also exhibit calorimetric cooperativity. This is
evident from their reported heat capacity scans
although $\Delta H_{\rm vH}/\Delta H_{\rm cal}$ ratios were not 
computed in these works.
\\

The role of local interactions is addressed
by a different coarse-grained model with G\=o-like 
(through-space) contact interactions but very little local (through-bond) 
preference for the CI2 native structure. The setup of this 
``contact-dominant'' model is similar to that of the NCS2 
without-solvation-SSR model: It has the same virtual bond strength 
($K_r=100\epsilon$), but the local native preference 
is weakened by a factor of 20, i.e., $K_\theta=0.5\epsilon$, 
$K_\phi^{(1)}=0.05\epsilon$, and $K_\phi^{(2)}=0.025\epsilon$.
Folding in this model is clearly non-two-state. In our simulation
of this contact-dominant model, conformations very close to the 
target native structure were observed but $Q=1$ was not achieved.\footnote
{
We have also studied similar ``contact-only'' models with 
$K_\theta=K_\phi^{(1)}=K_\phi^{(2)}=0$ in the same without-solvation-SSR 
setup as well as in the (full LJ) without-solvation formulation. 
These models have even bigger difficulties reaching conformations 
with $Q\approx 1$ than the contact-dominant model.
}
A numerical estimate of this model's heat capacity function 
was obtained from Langevin dynamics simulation near the transition 
midpoint. Figure~5 shows that it has a double hump, which is clearly 
dissimilar to the single-peak heat capacity scans of two-state proteins 
such as CI2.$^{72,95}$ Moreover, near this model's temperature for the 
peak heat capacity value, the distribution of $Q$ has only a single 
population maximum rather than being bimodal (data not shown). Indeed,
a few highest and lowest $Q$ values were so improbable that they were 
not sampled. These thermodynamically non-cooperative features are 
reflected by an exceedingly low van't Hoff over calorimetric
enthalpy ratio of $\kappa_2^{\rm (s)}=0.33$.
\\

One may conceivably argue from the ``energetic vs. topological 
frustration'' perspective$^{28,32}$ that energetic frustration
has already been eliminated in the contact-dominant model because its 
potential favors native contacts, disfavors nonnative contacts, and even 
slightly favors native bond angles and torsion angles. Yet the 
contact-dominant model's thermodynamics is not proteinlike.
The non-cooperative behavior of this particular contact-dominant 
model might have been exasperated by the exclusion of $(i,i+3)$ contacts
in its formulation (see equation~1). Nonetheless,
the present result echoes several recent 
findings of less-than-proteinlike thermodynamic cooperativity in continuum 
models with G\=o-like contact interactions but without local conformational 
preferences. These model studies include coarse-grained and all-atom 
discontinuous molecular dynamics models$^{41,45}$ as well as a 
self-consistent field theory.$^{37}$ On the other hand,
some three-dimensional lattice ``contact-only'' G\=o models are 
thermodynamically cooperative,$^{47}$ probably because of default lattice 
restrictions on bond angles and torsion angles. However, in continuous 
space, the ``negative design'' afforded by G\=o-like contact interactions 
alone are apparently insufficient for proteinlike thermodynamics. 
Indeed, a protein sequence's ability to fold to a unique structure may be 
partially encoded in local signals.$^{96,97}$ 
Proteinlike behavior requires minimization of energetic 
frustration of the target native structure as well as enhanced frustration 
in the competing nonnative conformations.$^{98}$ A comparison between the 
contact-dominant model and the other models with local native propensities 
in Figure~5 suggests that an interplay between local 
conformational preference and nonlocal compactification 
forces$^{47,48,67,68,99,100}$ are necessary for proteinlike thermodynamic 
cooperativity. For this conclusion to be properly interpreted, we hasten
to add that structural details of sidechain packing, hydrogen bonding, as well 
as general non-native-centric physical restrictions on bond angles and torsion 
angles (as in standard non-G\=o-like force fields) have not been taken 
into account in the present coarse-grained (residue-based) contact-dominant 
model. But these effects are operative in real proteins. Clearly, these 
interactions must be part of the physical basis of any local 
propensity$^{101}$ for the native fold in a more complete all-atom description.
\\


$\null$

\centerline{\bf KINETIC COOPERATIVITY}

$\null$

\noindent
{\large\it Sharp Kinetic Transitions Between Two Thermodynamic States}

Folding kinetics in explicit-chain G\=o-like models have been investigated 
using equilibrium sampling in conjunction with
free energy profile analyses$^{32}$ as well as direct dynamics 
simulations.$^{35}$ Here, around their respective transition midpoints,
all six native-centric CI2
models --- NCS1 or NCS2, with or without solvation --- have
kinetic characteristics consistent with their thermodynamic two-state 
cooperativity. Figure~6a and b show that the kinetic transitions between 
the native and denatured ensembles are sudden and sharp.
Figure~6c and d show that the distributions of potential energy and $Q$ 
are bimodally well separated into native and denatured regions, and 
the correlation between potential energy and $Q$ is generally linear.
A consistency check has also been made using Figure~6c, which provided an 
average kinetic energy of $78.9$. Equating this with $3NT/2$ 
for $N=64$ (equipartition theorem) yields $T=0.8219$, which is 
essentially identical to the input simulation temperature of $T=0.82$, 
as it should. Figure~6c and d further indicate that after the
initiation of folding around the transition midpoint,
pre-equilibration of the denatured ensemble 
is rapid relative to the folding time scale.
\\

\noindent
{\large\it Chevron Plots: Matching Kinetics with Thermodynamics?}

Bearing in mind that protein thermodynamic 
cooperativity is necessary but not sufficient for simple two-state 
folding/unfolding kinetics,$^{68}$ we proceed to
evaluate model predictions 
against experimental stability curves and chevron plots. To do so,
we determine model 
folding and unfolding rates using direct dynamics simulations over extensive 
ranges of native stability by varying the interaction parameter $\epsilon$ 
at constant temperature.
Since the simulated kinetics are essentially 
single-exponential (see below), the folding or unfolding rate may be taken to 
be approximately the reciprocal of the corresponding mean first 
passage time (MFPT). The natural logarithms of the rates are plotted as 
functions of $\epsilon$ in Figures~7 -- 9. Inasmuch as it was computationally 
feasible, first passage times (FPTs) of a large number of trajectories 
were used to provide reliable estimates of MFPTs (Tables~1 -- 3). 
As one of us has argued,$^{98,102}$ the variation of $\epsilon$ may 
serve as a tentative model for varying denaturant concentration, 
though the detailed physics of how the effects of chemical 
denaturants should be incorporated into coarse-grained protein models is 
a subject of ongoing research.$^{103-106}$ Here we view the upper panels of 
Figures~7 -- 9 as model equivalences of chevron plots.
\\

Native stability curves of the models as functions of $\epsilon$ are 
plotted in the lower panels of Figures~7 -- 9. They show that 
the free energy of unfolding between the native minimum and low-$Q$ open 
conformations are approximately linear in $\epsilon$ (upper solid
and dashed curves). These quasi-linear stability curves estimated 
from simulation data around the transition midpoint correspond to
those obtained experimentally by empirical linear extrapolation
from directly measured data around the transition region.$^{107}$ 
In contrast, the free energy difference 
between the native minimum and a denatured-state ensemble encompassing 
low-$Q$ as well as intermediate-$Q$ conformations (lower solid and dashed 
curves) is nonlinear in $\epsilon$, similar to that
observed in previously lattice model studies.$^{67,68}$ 
This is an expected feature$^{67,68}$ intimately connected to 
the multiple-conformation nature of the native state,$^{47,68}$ and is 
consistent with recent native-state hydrogen exchange experiments.$^{107-110}$ 
These characteristics of native stability underscore the fact that the 
operational definition of calorimetric two-state behavior (see above) 
does not$^{67}$ necessarily imply that 
all denatured conformations have the same stability. Even for calorimetrically 
two-state proteins under native conditions well below the 
global folding/unfolding transition midpoint, the population of partially 
unfolded conformations$^{67,68,107,111}$ can sometimes be non-negligible 
as long as it does not exceed a certain threshold.$^{67,68}$
\\

Figures~7 -- 9 show that the transition midpoints determined by
thermodynamics and kinetics are quite close, with only minor discrepancies.
The discrepancies for NCS1 models appear to be slightly larger in 
Figures~8 and 9.  This is probably related to the high-free-energy minima 
in the transition regions of the corresponding NCS1 free energy profiles 
(Figure~4b, c). More surprisingly, however, 
is that even with their native-centric potentials, all six models fail to 
produce the type of simple two-state folding/unfolding kinetics observed 
experimentally for CI2$^{95}$ and many other small single-domain 
proteins.$^{7}$ The operational definition$^{95,112}$ 
for simple two-state folding/unfolding kinetics requires that the 
logarithmic folding and unfolding rates under constant temperature be 
approximately linear in
native stability, and that the natural logarithm of the directly 
measured and linearly extrapolated (folding rate)/(unfolding rate) ratio 
as a function of denaturant concentration
matches the directly measured and
linearly extrapolated$^{95,107}$ equilibrium free energy of unfolding in 
units of $k_{\rm B}T$. Here, the dashed-dotted V-shapes in the upper panels 
of Figures~7 -- 9 show that as $-\epsilon/k_{\rm B}T$ is changed
at constant $T$ from the 
transition midpoint towards either more native or more denaturing conditions, 
the respective trends of increase in simulated folding or unfolding rate fall 
short of this requirement for the kinetics to be simple two-state. Instead,
our models' behavior is more akin to proteins that exhibit chevron rollovers, 
such as ribonuclease A$^{113}$ and barnase,$^{114}$ whose kinetics are 
operationally referred to as non-two-state.$^{68,95,113,114}$ Comparisons
with experimental chevron plots have not been made in other studies
of continuum G\=o models, but the reported 
results indicate that they also do not predict 
simple two-state chevron behavior (see, e.g., Figure~2 of ref.~34).
\\

The four without-solvation and without-solvation-SSR models in Figures~7 
and 8 show a clear rollover in both the folding and unfolding arms of 
their chevron plots. Reflecting the lower barriers along their free 
energy profiles (Figure~4), kinetics are generally faster for the 
without-solvation and without-solvation-SSR than the corresponding 
with-solvation models (Figure~9). For the with-solvation models, the 
rate at a given $-\epsilon/k_{\rm B}T$ is substantially slower for NCS2 
than that for NCS1. This trend is consistent with NCS2's 
much higher free energy barrier in the transition region (Figure~4c). Most 
remarkably, comparing Figures~7, 8 against Figure~9 demonstrates a dramatic 
impact of desolvation barriers on the folding/unfolding kinetics. In contrast 
to the chevron plots with significant curvatures for the without-solvation 
and without-solvation-SSR models, both the 
folding and unfolding arms of the chevron plots are quasi-linear for the 
with-solvation models. It is reassuring that the with-solvation models 
are more proteinlike in this respect.$^{113,114}$ Nevertheless,
their deviations from simple two-state kinetics are huge: 
The slopes of the simulated chevron plots are only approximately
1/5 that required for simple two-state kinetics 
(c.f. the V-shape in the upper panel of Figure~9). Therefore, 
the conclusion that these models' kinetics correspond to those operationally
referred to as non-two-state should be reliable. This is because
possible numerical uncertainties in the estimation of stability curves
by histogram techniques (lower panels of Figure~9) are not
likely to cause a factor-of-five discrepancy. Interestingly, similar
mismatches between extrapolated chevron plots and direct native stability 
measurements, albeit to a lesser degree, have also been observed for
real proteins.$^{93}$
\\

\noindent
{\large\it Single-Exponential Relaxation}

Experimentally, kinetic relaxation of many small single-domain proteins$^{7}$ 
such as CI2$^{72,95}$ and some apparently non-two-state proteins with chevron 
rollovers$^{113,114}$ are found to be essentially single-exponential. Therefore,
it is of interest to ascertain whether the present models embody this 
hallmark, even though they are not kinetically simple two-state.
For this purpose, we examine the distributions of first passage 
times (FPTs, as defined in Figures~7 -- 9). Let $P(t)dt$ be the
probability for the FPT of a given kinetic process to lie within a range 
$dt$ around time $t$. If the relaxation is single-exponential,
\begin{equation}
\int_{t_0}^t dt^\prime\; P(t^\prime) = 1 - e^{-k(t-t_0)} \; ,
\end{equation}
where $k$ is the kinetic rate, and $t_0\ge 0$ is a minimum FPT to take into 
consideration the finite time needed for pre-equilibration after 
initiation of the kinetic process at $t=0$.
It follows that
\begin{equation}
{\rm MFPT} = \int_{t_0}^t dt^\prime\; t^\prime
P(t^\prime) = t_0 + {\frac 1 k} \; .
\end{equation}
To assess whether a given FPT distribution conforms to this 
description, a quantity $P(t)\Delta t$ is computed by binning FPTs into 
time slots$^{115}$ of size $\Delta t$. If the kinetic process
is single-exponential,
\begin{equation}
\ln [P(t)\Delta t] = \biggl\{ 
\ln \biggl({\frac {\Delta t} {{\rm MFPT}-t_0}} \biggr ) 
+ {\frac {t_0} {{\rm MFPT}-t_0}} \biggr \} 
- {\frac {t} {{\rm MFPT}-t_0}} \; ,
\end{equation}
i.e., $\ln [P(t)\Delta t]$ versus $t$ should be
a straight line with slope $=$ $-({\rm MFPT}-t_0)^{-1}$.
\\

Figure~10a shows that even under strongly native conditions concomitant 
with a significant chevron rollover, the NCS2 without-solvation-SSR 
model has approximately single-exponential relaxation. 
This behavior echoes that of a recent four-helix-bundle lattice 
model$^{68}$ (Figure~10b). Consistent with equation~6, a comparison between 
the filled and open circles in Figure~10a indicates that while changing 
the bin size $\Delta t$ naturally changes the $\ln[P(t)\Delta t]$ 
values, reasonable variations in $\Delta t$ do not affect the slope of 
the $\ln[P(t)\Delta t]$ distribution. Figure~11 applies similar analyses 
to folding and unfolding in other models in the present study
under representative native and denaturing conditions.\footnote{
Rates in the chevron plots in Figures~7 -- 9 are computed by taking 
$t_0=0$. Our calculations indicate that using finite $t_0$s instead
of $t_0=0$ to determine the rates $k$ via equation~5 only leads to minimal 
modifications on the chevron plots (data not shown). The conclusions 
regarding rollovers and non-two-state kinetics remain unchanged.
}
Owing to computational limitations, the sample 
sizes for the FPT distributions are not very large, especially for 
the with-solvation models in Figure~11c. Consequently, a certain level 
of statistical uncertainties ensued. Nonetheless, Figure~11 shows that 
for all cases 
tested, our data is consistent with single-exponential relaxation.
As pointed out by Fersht,$^{92}$ the high-free-energy minima along
the NCS1 free energy profiles (Figure~4b, c) do not preclude apparent 
single-exponential kinetics. The viability of equation~6 for our models
is further buttressed by the relatively small differences between the 
slopes of the least-square-fitted lines in Figure~11 and the quantity 
$-({\rm MFPT}-t_0)^{-1}$, where $t_0$ is taken to be the minimum FPT 
encountered in the simulated trajectories of a given model:
For the models and their simulation conditions listed in the legend 
of Figure~11, and in the same order, 
$\{[10^6\times ({\rm MFPT}-t_0)^{-1}],[-10^6\times {\rm slope}]\}=$
$\{ 10.0, 10.6\}$, $\{ 6.13, 6.71\}$, $\{ 6.11, 6.53\}$, $\{ 7.28, 7.56\}$,
$\{ 28.6, 36.1\}$, $\{ 6.41, 7.02\}$, $\{ 19.5, 26.4\}$, $\{ 8.53, 9.36\}$,
$\{ 2.05, 2.08\}$, $\{ 0.504, 0.431\}$, $\{ 0.242, 0.189\}$, and
$\{ 0.199, 0.161\}$.
\\

The native-centric formulations in the present G\=o-like models lead to 
folding rates that are at least four orders of magnitdue faster than the 
experimental CI2 folding rates. At 25$^\circ$C and pH 6.3, the experimental 
CI2 folding rates at zero denaturant (native stability 
$\Delta G=12.0 k_{\rm B}T$) and the transition midpoint (in 3.92 M GdnHCl, 
native stability $\Delta G=0$) are, respectively, $47.8$ sec$^{-1}$ and 
$0.035$ sec$^{-1}$ (ref.~95). If we use the physical argument of Veitshans 
et al.$^{89}$ to identify the Langevin time unit $\delta t$ with a real
time scale of $\sim 10^{-14}$ sec, the folding rate of the NCS2 
with-solvation model in Figure~9 is $\sim 10^6$ sec$^{-1}$ at 
$\Delta G=12.0 k_{\rm B}T$ and $\sim 10^5$ sec$^{-1}$ at $\Delta G=0$. 
Corresponding folding rates of other models in Figures~7 -- 9 are even 
faster by approximately two orders of magnitude. Despite these discrepancies,
native-centric constructs do capture part of real protein energetics. This 
is evident from studies of extensive sets of real proteins using
explicit-chain G\=o models, wherein theoretically predicted folding$^{35}$ 
and relaxation$^{44}$ rates were found to correlate reasonably well with the 
experimental folding rates. However, it is noteworthy that the spread of 
these model-predicted rates among the set of proteins tested is apparently 
at least 1.5 -- 2 orders of magnitude narrower than the diversity of 
experimental folding rates.$^{35}$ (c.f. Figure~5 of ref.~44). This suggests 
that certain basic aspects of protein energetics are yet to be taken into 
account by common G\=o-like models. In a similar vein, the chevron rollovers 
in Figures~7 --9 represent a failure to account for the high degree of 
diversity in folding rates of a given protein under different native 
conditions. For real CI2, the folding rates at zero denaturant and
at the transition midpoint differ by three orders of magnitude. But
the G\=o-like models in Figures~7 --9 predict only one order of magnitude 
difference.
\\

\noindent
{\large\it Chevron Rollover: Stability-Dependent Front Factor?}

To better understand the chevron rollovers, Figure~12
applies a protocol we recently developed$^{68}$ to assess the 
models' conformity to the commonly employed transition state picture 
in interpreting protein folding experiments. Model data is now fitted 
to the expression 
\begin{equation}
{\rm rate} = F(\epsilon,T) 
\exp\biggl[-{\frac {\Delta G^\ddagger(\epsilon,T)} {k_B T}}\biggr] \; 
\end{equation}
for folding or unfolding rate, taken as $({\rm MFPT})^{-1}$ from 
the direct dynamics simulations. On the other side of the above equation, 
$\Delta G^\ddagger$ is an activation free energy determined soley by 
thermodynamic Boltzmann weights$^{68}$ using the method of Nymeyer 
et al.,$^{32,116}$ $F$ is the corresponding front factor.$^{2,38,39,68,98,117}$
Figure~12 shows that, in contrast to the usual stipulation$^{118}$ that
the front factors of small single-domain proteins such as CI2 are
essentially independent of intraprotein interaction strength and
native stability, the $F$ factors deduced from the present analysis are 
highly sensitive to $\epsilon$.
This implies that thermodynamic analyses of free energy profiles alone 
cannot predict the $\epsilon$-dependencies of the kinetic 
rates,$^{38,39,68,117}$ and the chevron rollovers are underpinned by
native-stability-dependent front factors in these models.$^{68}$
This hypothesis regarding the physical origin of chevron rollover
may soon be testable by single-molecule techniques.$^{119}$
In addition to the definitions for unfolded, transition, and folded 
regions in Figure~12, we analyzed several other physically 
reasonable alternate $Q$-based definitions for these states 
(data not shown).  Whereas the absolute value of $F$ varies 
somewhat, the overall trend of dependence on $\epsilon$ remains essentially 
unchanged. This resilience is similar to that observed in our previous 
analysis of the folding front factor of a 55mer lattice model (Figure~5 of
ref.~68). 
\\

Thus, for this key aspect of chevron behavior, the present native-centric 
models' kinetics clearly do not resemble
the simple two-state kinetics of CI2.$^{72,95}$ The
ramification of this finding is far reaching, as it bears on the 
basic energetics of protein folding (see Discussion below). As they stand,
the apparently non-two-state kinetics of these physical self-contained polymer 
models$^{18}$ also shed light on the folding of other proteins that 
exhibit similar chevron rollovers as well.$^{93,113,114}$ 
To date, rationalizations 
of chevron rollovers include deadtime intermediates,$^{120}$ specific kinetic 
traps,$^{98,102,121}$ peak-shifting on complex free energy profiles,$^{93,122}$ 
burst phase continuum,$^{123}$ and internal friction as manifested by 
front factors that depend on native stability (ref.~68 and discussion
therein). These perspectives are not necessarily mutually exclusive.
For example, internal friction may arise from kinetic trapping mechanisms 
(H. K.  \& H. S. C., in preparation). In any event, chevron rollover is an 
unequivocal prediction of the present models, irrespective of whether 
$Q$ or other folding reaction coordinates are used 
for the transition state analysis (see, e.g., ref.~124). 
Figure~12d shows that the folding front factor 
decreases with more native conditions, and the unfolding front factor 
also decreases with more denaturing conditions. In short, there appears to be 
an aversion to speed in these models' energetics.
We tentatively attribute the slowing down in 
these models to a possible combination of effects of 
internal friction (conformational search problems compounded by more native 
conditions)$^{68}$ and external friction (implicit solvent viscosity). 
The origins of these effects remain to be better elucidated. For 
example, in some modeling situations,$^{68}$ folding-arm rollovers are 
related to the onset of downhill folding.$^{125,126}$ The chevron 
rollovers in the folding and unfolding arms of the NCS2 without-solvation 
model may be similarly related to downhill scenarios 
(see, e.g., the $\epsilon=0.90$ and $\epsilon=0.70$ profiles in Figure~12a).
At least for the NCS2 with-solvation model in Figure~6, 
the fact that no deadtime intermediate was 
observed during our simulation suggests that such a mechanism is
not necessary for chevron rollovers.$^{68,93}$ In this example, chevron
rollover emerges as a kinetic front-factor effect.
\\
 

$\null$

\centerline{\bf DISCUSSION}

$\null$

We have compared two different native contact sets, and three different 
formulations of G\=o-like interactions with and without 
desolvation barriers. The predictions of these native-centric models were 
evaluated against generic thermodynamic and kinetic properties of small 
single-domain proteins that these models were designed to mimic in the first
place.  We learnt several lessons.  First, proteinlike thermodynamic 
cooperativity requires nonlocal contact-like interactions acting
in concert with local conformational favorabilities for the native 
fold$^{47,48,67,68}$ (Figure~5). Second, some basic predictions of 
native-centric models, such as the free energy profiles in Figure~4, are 
significantly dependent on the native contact set and interaction scheme 
used, even if the choice is made among physically reasonable definitions.
A recent study$^{38,39}$ shows also that free energy 
profiles of native-centric models are sensitive to the chain's 
presumed persistence length and 
energetic barriers to bond rotations.$^{98,102}$ 
Third, we found that pairwise desolvation barriers in native-centric 
models could lead to some proteinlike properties such as a higher free energy 
barrier separating the native and denatured states (c.f. Figure~4a, b and c) 
as well as more linear chevron plots (c.f. Figures~7, 8 and 9). These 
predictions are encouraging as they provide insight into 
corresponding features in real proteins. 
Fundamentally, however, the kinetics of all present native-centric models 
for CI2 do not resemble that of real CI2. The models with pairwise 
desolvation barriers, like those without, are kinetically non-two-state in
the operational sense that they have large chevron rollovers. 
\\

Fourth, the significant differences between the predictions of with- 
and without-solvation models underscore the importance of proper
accounting for the energetic cost of water expulsion in protein 
folding models, and that caution should be used when interpreting 
results obtained from effective potentials that do not have 
desolvation barriers.$^{83,127}$ The barrier height in the present 
with-solvation models simulated at $T=0.82$ is
$0.24\; \epsilon\;  k_{\rm B}T$. 
For real proteins, the desolvation barrier heights encountered by the 
polypeptide chain as a part of the potential of mean force are expected
to be sensitive to temperature. Thus, the present results 
should also bear on explicit-solvent 
unfolding simulations at high temperatures and the degree of dependencies 
of protein folding mechanisms on temperature.$^{128}$ Of relevance here
is the model system of a pair of methanes in water. 
Recent Monte Carlo simulations in the TIP4P water 
model indicate that their desolvation barrier is reduced from approximately 
$0.16$ to $0.12$ kcal/mol ($0.27 k_{\rm B}T$ to $0.16 k_{\rm B}T$) 
when temperature is increased from 298K to 368K under atmospheric 
pressure.$^{84}$ 
Under typical high-temperature unfolding conditions of 
498K and a water density of 0.829 gm/ml,$^{128}$ 
the desolvation barrier height 
is further reduced to $\approx 0.05$ kcal/mol or $0.05 k_{\rm B}T$
(Figure~16.3 in ref.~18; S. Shimizu and H. S. C., personal communication). 
\\

In a broader perspective, solvent-mediated interactions 
are known to be intrinsically 
pairwise nonadditive,$^{76,83}$ and the collapse of a hydrophobic chain may 
involve large length-scale dynamic effects.$^{129}$ In this light, that the 
pairwise desolvation barriers here fail to produce simple
two-state chevron plots is 
not too surprising. Indeed, recent explicit-water simulations show that 
the sign of heat capacity of the free energy
barrier against folding is opposite to that against
the association of a pair of methane molecules.$^{84,85,106}$ Considerations
of a three-methane model system further indicate that the height of 
desolvation barrier is clearly nonadditive, and the sign and magnitude of 
this nonadditivty is dependent upon the configuration of the
nonpolar solutes involved.$^{83}$ Hence, solvation effects beyond
the pairwise formulation considered here are likely needed to
account for simple two-state protein folding/unfolding kinetics.
\\

In summary, the present findings imply that the actual solvent-mediated 
interactions in real proteins are much more specific and well-designed 
than one would otherwise posit. In short, real proteins are more
cooperative than common G\=o-like models with pairwise additive
interactions. Nonetheless, recent 
innovations in native-centric modeling have been immensely valuable. As 
discussed above, they do capture part of the essential physics. Many deep 
insight would not have been gained without them 
(see, e.g., refs.~1, 29, 33). 
But, at the same time, the limitations of common G\=o-like 
chain models$^{67,68}$ may be more basic than previously 
appreciated. The present analysis implies that more 
proteinlike interaction schemes are yet to be discovered. Every
model considered here except the contact-dominant variety
can fold to the CI2 native structure. Qualitatively, 
the free energy profiles of the 
NCS2 models fit the expectation for that of small single-domain proteins as 
well.  Yet their kinetics are fundamentally different from that of 
CI2. Thus, a protein model's ability to fold to one single target structure
does not guarantee the adequacy of its energetics; and the microscopic
origin of simple two-state folding/unfolding kinetics remains to be 
elucidated. Our effort to address some of these questions is underway.
Apparently, chevron rollovers can be essentially eliminated in more 
cooperative chain models with added energetic favorabilities 
for the ground-state and near-ground-state structures beyond that 
provided by the additive schemes in common G\=o models. These results will 
be presented in a subsequent report (H. K. \& H. S. C., in preparation). 
In the ongoing quest for 
a better understanding of protein energetics through the design and 
interpretation of novel physical models, proteinlike statistical mechanics 
properties such as calorimetric two-state 
cooperativity$^{41,47,48,67,68,130,131}$ 
and simple two-state chevron behavior$^{68}$ should be useful
as stringent but necessary modeling constraints.
\\

$\null$

\noindent
{\Large Acknowledgments}

We thank Yawen Bai, Margaret Cheung, Cecilia Clementi, Ken Dill, 
Angel Garc{\'\i}a, Chinlin Guo, Carol Hall, Anders Irb\"ack, Sophie Jackson, 
John Karanicolas, Bob Matthews, Cristian Micheletti, Hugh Nymeyer, 
Mikael Oliveberg, Jos\'e Onuchic, Kevin Plaxco, John Portman, Eugene
Shakhnovich, Joan-Emma Shea, Seishi Shimizu, Kim Sneppen, Dev Thirumalai, 
Michele Vendruscolo, Peter Wolynes, and Yaoqi Zhou for helpful discussions 
during the period in which these ideas were developed. 
This work was partially supported by the Canadian Institutes of 
Health Research (CIHR grant no. MOP-15323), a Premier's Research 
Excellence Award from the Province of Ontario, and the Ontario Centre 
for Genomic Computing at the Hospital for Sick Children in Toronto.
H. S. C. is a Canada Research Chair in Biochemistry.

\vfill\eject


\noindent
{\large\bf References}

\kern -1.5cm

\vfill\eject


\centerline{\large \bf Table 1}
\vskip .2 in

\begin{center}
\begin{tabular}{|r|r|r|r|r||r|r|r|r|r|}
\hline
\multicolumn{5}{|c||}{{\bf unfolding}} & 
\multicolumn{5}{|c|}{{\bf folding}} \\
\hline
\multicolumn{1}{|c|}{$\null$}&\multicolumn{2}{|c|}{NCS1}&
\multicolumn{2}{|c||}{NCS2}&
\multicolumn{1}{|c|}{$\null$}&\multicolumn{2}{|c|}{NCS1}&
\multicolumn{2}{|c|}{NCS2}\\
\cline{2-5} \cline{7-10}
\multicolumn{1}{|c|}{$\epsilon$} & 
\multicolumn{1}{|c|}{\footnotesize MFPT/$10^5$} &
\multicolumn{1}{|c|}{$N_t$} &
\multicolumn{1}{|c|}{\footnotesize MFPT/$10^5$} &
\multicolumn{1}{|c||}{$N_t$} &
\multicolumn{1}{|c|}{$\epsilon$} & 
\multicolumn{1}{|c|}{\footnotesize MFPT/$10^5$} &
\multicolumn{1}{|c|}{$N_t$} &
\multicolumn{1}{|c|}{\footnotesize MFPT/$10^5$} &
\multicolumn{1}{|c|}{$N_t$} \\
\hline \hline
0.60 & 0.1734& 100& 0.1763& 100& 1.00&$0.6905$& 100&$0.9250$& 100\\
0.65 & 0.2472& 100& 0.2547& 100& 0.95&$0.9534$& 100&$1.1668$& 100\\
0.70 & 0.3825& 100& 0.4212& 100& 0.90&$1.4646$& 100&$1.3102$& 100\\
0.75 & 0.7807& 100& 0.8950& 100& 0.89&$1.5062$& 100&$1.3865$& 100\\
0.77 & 1.2540&1100& 1.9052&1100& 0.88&$1.8175$&1100&$1.5577$&1100\\
0.78 & 1.7684& 100& 2.8292& 100& 0.87&$2.1760$& 100&$2.0039$& 100\\
0.79 & 2.4983& 100& 4.7346& 100& 0.86&$2.2865$& 100&$2.3725$& 100\\
0.80 & 2.3120& 100& 8.0901& 100& 0.85&$3.0018$& 100&$2.7999$& 100\\
0.82 & 3.4755& 100& 42.737& 100& 0.84&$4.2971$& 100&$4.0108$& 100\\
---  & ---   & ---& ---   & ---& 0.83&$4.6969$& 100&$5.1892$& 100\\
---  & ---   & ---& ---   & ---& 0.82&$7.9260$& 100&$7.3979$& 100\\
---  & ---   & ---& ---   & ---& 0.80&$20.278$& 100&$17.844$& 100\\
\hline
\end{tabular}
\end{center}
\vskip .15 in

{\noindent {{\bf Table~1.}}} $\quad$ 
Number of trajectories $N_t$ used in the present study to determine
the MFPT of folding and unfolding for the
NCS1 and NCS2 without-solvation models ($T=0.82$, Figure~7).
Each MFPT listed is the average (arithmetic mean) over $N_t$ first
passage times for the given interaction strength $\epsilon$. Time is
measured from the start of a given simulation at $t=0$ in units of
$\delta t$ (see text).

\vfill\eject


\centerline{\large \bf Table 2}
\vskip .2 in

\begin{center}
\begin{tabular}{|r|r|r|r|r||r|r|r|r|r|}
\hline
\multicolumn{5}{|c||}{{\bf unfolding}} & 
\multicolumn{5}{|c|}{{\bf folding}} \\
\hline
\multicolumn{1}{|c|}{$\null$}&\multicolumn{2}{|c|}{NCS1}&
\multicolumn{2}{|c||}{NCS2}&
\multicolumn{1}{|c|}{$\null$}&\multicolumn{2}{|c|}{NCS1}&
\multicolumn{2}{|c|}{NCS2}\\
\cline{2-5} \cline{7-10}
\multicolumn{1}{|c|}{$\epsilon$} & 
\multicolumn{1}{|c|}{\footnotesize MFPT/$10^5$} &
\multicolumn{1}{|c|}{$N_t$} &
\multicolumn{1}{|c|}{\footnotesize MFPT/$10^5$} &
\multicolumn{1}{|c||}{$N_t$} &
\multicolumn{1}{|c|}{$\epsilon$} & 
\multicolumn{1}{|c|}{\footnotesize MFPT/$10^5$} &
\multicolumn{1}{|c|}{$N_t$} &
\multicolumn{1}{|c|}{\footnotesize MFPT/$10^5$} &
\multicolumn{1}{|c|}{$N_t$} \\
\hline \hline
0.70 & 0.1804&  100& 0.1856& 100& 1.30&$1.0080$&100&$0.9121$&100\\
0.75 & 0.2450& 100& 0.2730& 100& 1.25&$1.1324$&100&$1.0924$&100\\
0.80 & 0.3362& 100& 0.3759& 100& 1.20&$1.4063$&100&$1.0754$&100\\
0.83 &  0.4489& 100& 0.5304& 100& 1.18&$1.4840$&100&$1.2380$&100\\
0.85 &  0.5653&1100& 0.6950&1940&1.15&$1.7905$&1100&$1.6733$&1097\\
0.88 &  0.9161& 100& 1.2724& 100& 1.13&$2.3409$&100&$1.6971$&100\\
0.90 & 1.3364& 100& 2.3473&100& 1.10&$2.8513$&100&$2.2875$&100\\
0.93 & 2.3120& 100& 6.5846&100& 1.08&$4.0404$&100&$3.0059$&100\\
0.95& 4.7098& 100& 15.767&100&1.05&$6.8365$&100&$4.8426$&100\\
0.97 & 9.4389& 75 & 40.261&45&1.03&$10.385$&100&$9.6890$&100\\
0.98&16.268& 91 & 42.673&20&1.00&$43.174$&52&$25.440$&60\\
0.99 &   ---  & ---& 71.042&15&0.98&---&---&110.78&24\\
1.00&51.148& 18 &262.14&7& 0.97&---&---&349.46&14\\
\hline
\end{tabular}
\end{center}
\vskip .15 in

{\noindent {{\bf Table~2.}}} $\quad$ 
Same as Table 1, but for the without-solvation-SSR models ($T=0.64$, Figure~8).

\vfill\eject


\centerline{\large \bf Table 3}
\vskip .2 in

\begin{center}
\begin{tabular}{|r|r|r|r|r||r|r|r|r|r|}
\hline
\multicolumn{5}{|c||}{{\bf unfolding}} & 
\multicolumn{5}{|c|}{{\bf folding}} \\
\hline
\multicolumn{1}{|c|}{$\null$}&\multicolumn{2}{|c|}{NCS1}&
\multicolumn{2}{|c||}{NCS2}&
\multicolumn{1}{|c|}{$\null$}&\multicolumn{2}{|c|}{NCS1}&
\multicolumn{2}{|c|}{NCS2}\\
\cline{2-5} \cline{7-10}
\multicolumn{1}{|c|}{$\epsilon$} & 
\multicolumn{1}{|c|}{\footnotesize MFPT/$10^5$} &
\multicolumn{1}{|c|}{$N_t$} &
\multicolumn{1}{|c|}{\footnotesize MFPT/$10^5$} &
\multicolumn{1}{|c||}{$N_t$} &
\multicolumn{1}{|c|}{$\epsilon$} & 
\multicolumn{1}{|c|}{\footnotesize MFPT/$10^5$} &
\multicolumn{1}{|c|}{$N_t$} &
\multicolumn{1}{|c|}{\footnotesize MFPT/$10^5$} &
\multicolumn{1}{|c|}{$N_t$} \\
\hline \hline
0.40&0.5559&600&0.6549&500& 1.50&1.2196&100&2.1393&150\\
0.50&0.8419&126&1.5782&137& 1.40&2.4237&108&3.5297&176\\
0.60&1.6697&111&7.2719&121& 1.30&3.9479&117&5.5725&176\\
0.65&2.7900&103&17.122&50&  1.25&5.4452&107&8.7000&85\\
0.70&5.4940&1100&44.590&205&1.20&6.8649&112&11.766&50\\
0.75&8.4002&120&104.18&34&  1.18&8.7936&119&---&---\\
0.80&15.463&91&293.56&26&   1.15&10.823&52&30.141&42\\
0.83&26.101&51&---&---&     1.10&19.633&427&48.000&205\\
0.85&42.512&74&722.32&15&   1.08&20.336&36&---&---\\
0.90&119.01&32&1224.2&6&    1.05&41.682&63&143.75&24\\
0.92&---&---&1922.0&4&      1.03&43.365&56&166.15&20\\
0.93&168.05&24&---&---&     1.00&75.048&52&330.35&28\\
0.95&216.32&37&---&---&     0.97&93.742&40&444.98&7\\
    &      &  &   &   &     0.95&132.12&37&783.56&6\\
    &      &  &   &   &     0.90&---&---&1519.4&3\\
\hline
\end{tabular}
\end{center}
\vskip .15 in

{\noindent {{\bf Table~3.}}} $\quad$ 
Same as Table 1, but for the with-solvation models ($T=0.82$, Figure~9).

\vfill\eject


$\null$
\vskip -1.4cm

\noindent
{\large\bf Figure Captions}\\

\noindent
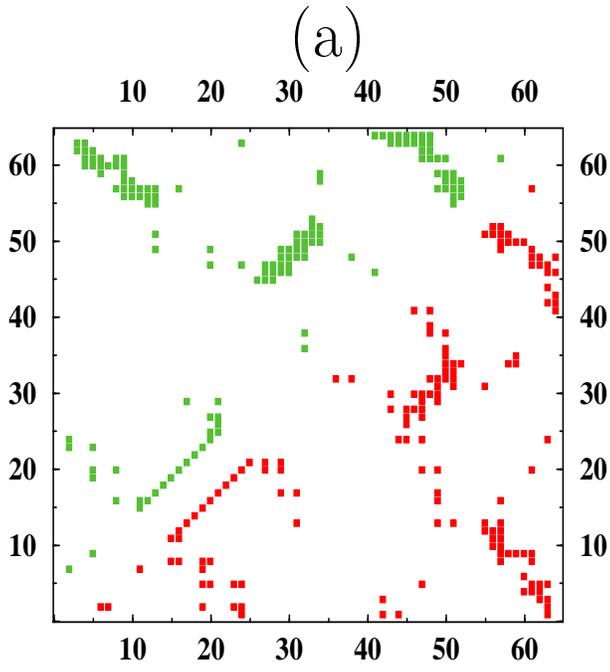
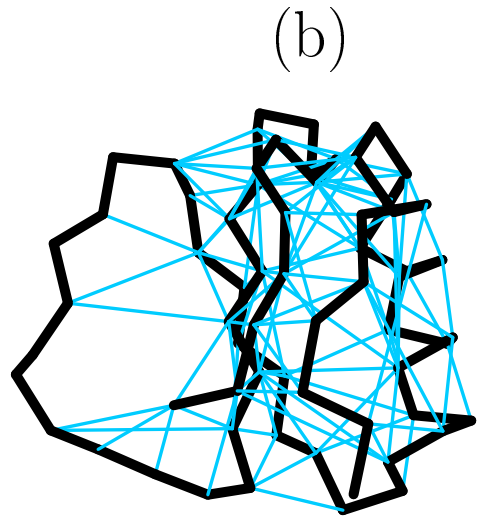
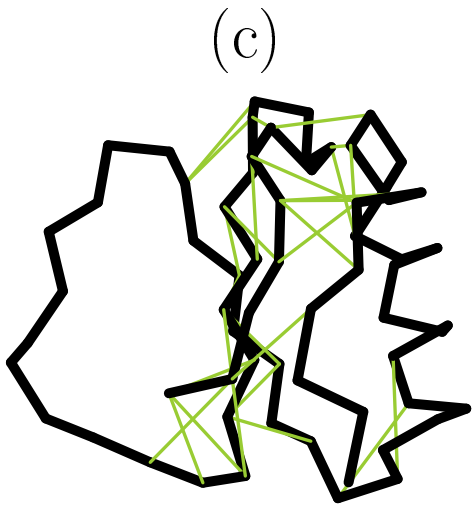
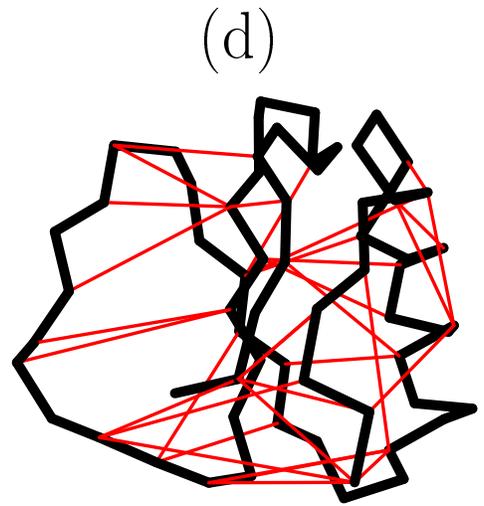
{\bf Figure 1.} $\quad$
Native contact maps of the 64-residue truncated form of chymotrypsin 
inhibitor 2 (2ci2) used in the present investigation. (a) Contact 
maps for the native contact sets NCS1 (green dots) and NCS2 (red dots) 
as defined in the text. Numbering of amino acids in these maps is 
initialized at residue 20 of the full-length 83-residue CI2; i.e., residue 
1 in (a) corresponds to Leu 20 in the untruncated protein. (b, c, d) 
Similarities and differences between native contact maps. Contact pairs 
are indicated by color lines joining C$_\alpha$ positions along the 
backbone (black trace). (b) Contacts shared by NCS1 and NCS2. (c)
Contacts in NCS1 but not in NCS2. (d) Contacts in NCS2 but not in NCS1.
\\

\noindent
{\bf Figure 2.} $\quad$
Different native contact definitions.
Amino acid numbering here corresponds to that of the full-length CI2, 
i.e., numbering in this figure equals to that in Fig.~1a plus 19.
The red contact belongs to NCS2 but not NCS1. This pair of residues
has a C$_\alpha$--C$_\alpha$ distance of 11.24 \AA, with closest atomic 
separation between the residues equals 4.3 \AA. 
The green contact belongs to NCS1 but not NCS2. The C$_\alpha$--C$_\alpha$ 
distance between this pair of residue is 5.36 \AA.
\\

\noindent
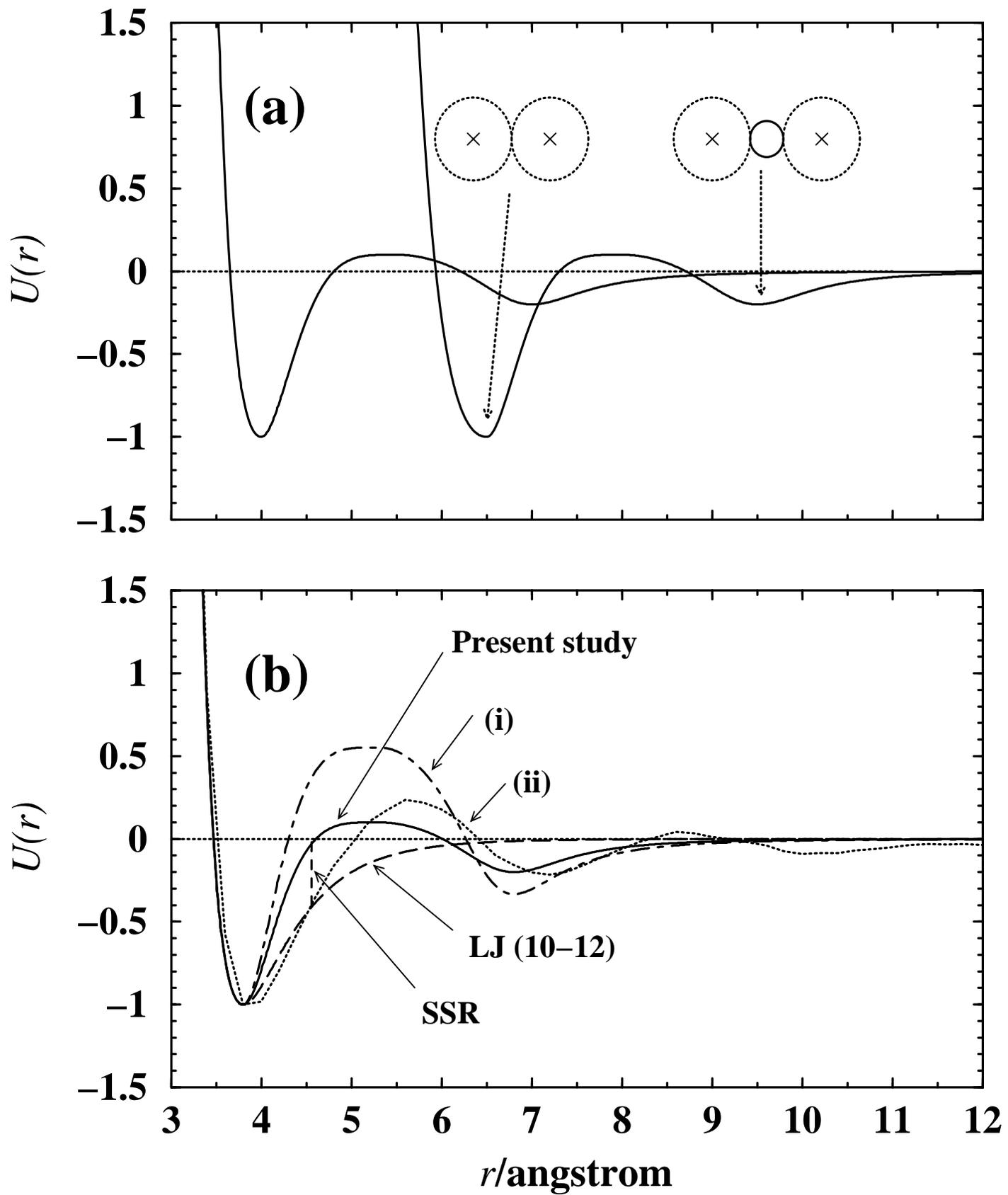
{\bf Figure 3.} $\quad$
(a) Model with-solvation interactions between two amino acid residues 
belonging to a given native contact pair in the present study 
(defined by either NCS1 or NCS2); $r$ is their C$_\alpha$--C$_\alpha$ 
separation. The potential energy $U(r)$, shown here and in part (b)
in units of $\epsilon$, depends also on the native C$_\alpha$--C$_\alpha$ 
distance $r^\prime$ of a given contact in the PDB structure. 
The $r^\prime$ values shown in this figure are only for illustrative 
purposes. They do not correspond to actual contacts in the present 
NCS1 or NCS2 models (see text).
$U(r)$ here is defined by the functional form of Cheung 
et al.$^{40}$ with $k=6$, $n=2$, $m=3$, $\epsilon^\prime=0.2\epsilon$, and 
$\epsilon^{\prime\prime}=0.1\epsilon$, where $k$, $n$, and $m$
parametrize the functional form for $r<r^\prime$, $r^\prime\le r< r^\dagger$ 
and $r\ge r^\dagger$, respectively (e.g., the excluded volume
repulsion $\sim r^{-2k}$, see equation on page~689 of ref.~40 for
details), $\epsilon^\prime$ is the depth of 
water-separated minimum, and $\epsilon^{\prime\prime}$ is the height of    
the desolvation peak. The two potential functions shown in (a) are 
for $r^\prime=4.0$ \AA~ (left) 
and $r^\prime=6.5$ \AA~ (right). The cartoons (for $r^\prime=6.5$ \AA)
illustrate contact and water-separated minima configurations,$^{40,83}$ 
where a water molecule is depicted as a solid circle of diameter $\approx
3$ \AA. (b) With-solvation native potential for 
$r^\prime=3.8$ \AA~ in the present study (as labeled) is compared with:
{\bf (i)} The $r^\prime=3.8$ \AA~ functional form of Cheung et al. with the 
same values for $k$, $n$, $m$, and $\epsilon$ as in (a), but with 
$\epsilon^\prime=\epsilon/3$ and $\epsilon^{\prime\prime}=5\epsilon/9$.
{\bf (ii)} The explict-water simulated methane-methane PMF at 25$^\circ$ C
under atmospheric pressure obtained by Shimizu and Chan,$^{83}$ in a unit 
such that the free energy at contact equals $-\epsilon=-1$. {\bf (LJ)} 
The $10$--$12$ Lennard-Jones without-solvation
potential $\epsilon[5(r^\prime/r)^{12}-6(r^\prime/r)^{10}]$ (as in 
equation~1) with $r^\prime=3.8$ \AA. {\bf (SSR)} The corresponding LJ cutoff 
at $1.2 r^\prime$ in without-solvation-SSR models.
\\

\noindent
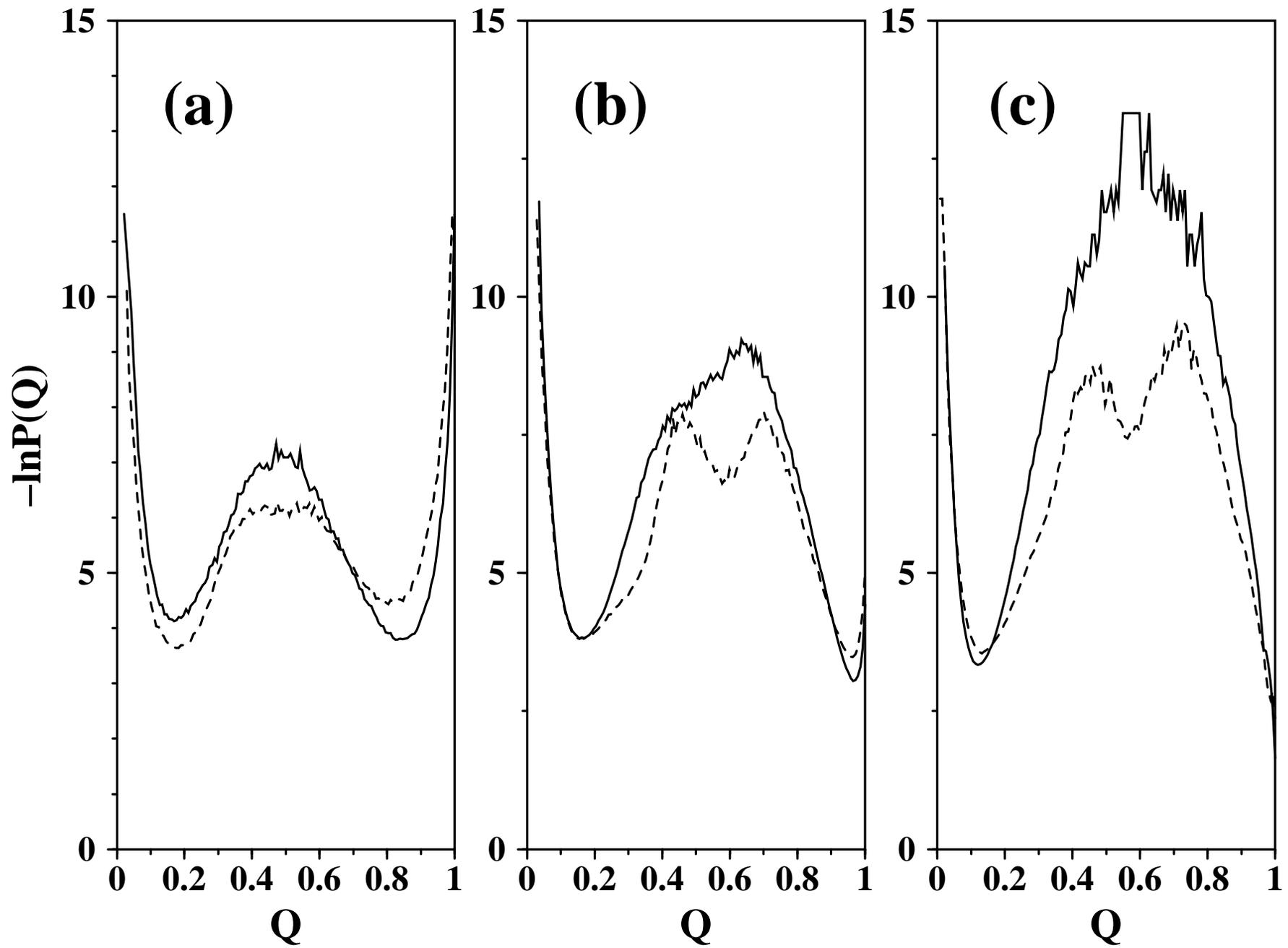
{\bf Figure 4.} $\quad$
Free energy profiles for NCS1 (dashed curves) and NCS2 (solid curves), 
using native contact potentials without (a, b) and with (c) desolvation 
barriers. (a) is for without-solvation models that use the full 
spatial range of the LJ terms whereas (b) is for 
without-solvation-SSR models with LJ cutoffs. The variable $Q$ is the number 
of native contacts in a conformation divided by the number of contacts in 
the native conformation of the given model. $P(Q)$ is the normalized 
population distribution over $Q$. The $-\ln P(Q)$ profiles 
are computed at each model's approximate transition midpoint:
(a) at $\epsilon/k{\rm _B} T$ $=$ $0.988$ and $0.988$ for the NCS1 and NCS2 
without-solvation models, (b) at $\epsilon/k{\rm _B} T$ $=$ $1.563$ 
and $1.547$ for the NCS1 and NCS2 without-solvation-SSR models, 
and (c) at $\epsilon/k_{\rm B} T$ $=$ $1.165$ and $1.098$ for the NCS1 and NCS2 
with-solvation models. 
For without-solvation models in (a) and (b), the condition for contact 
is $r\le 1.2 r^\prime$, as in ref.~32.
For with-solvation models in (c), a pair of residues is defined to be in 
contact when $r\le r^\dagger = (r^\prime+r^{\prime\prime})/2$, 
i.e., when the C$_\alpha$--C$_\alpha$ distance $r$ is within
the contact basin ($r$ not larger than that of the desolvation peak), as 
in ref.~40.
\\

\noindent
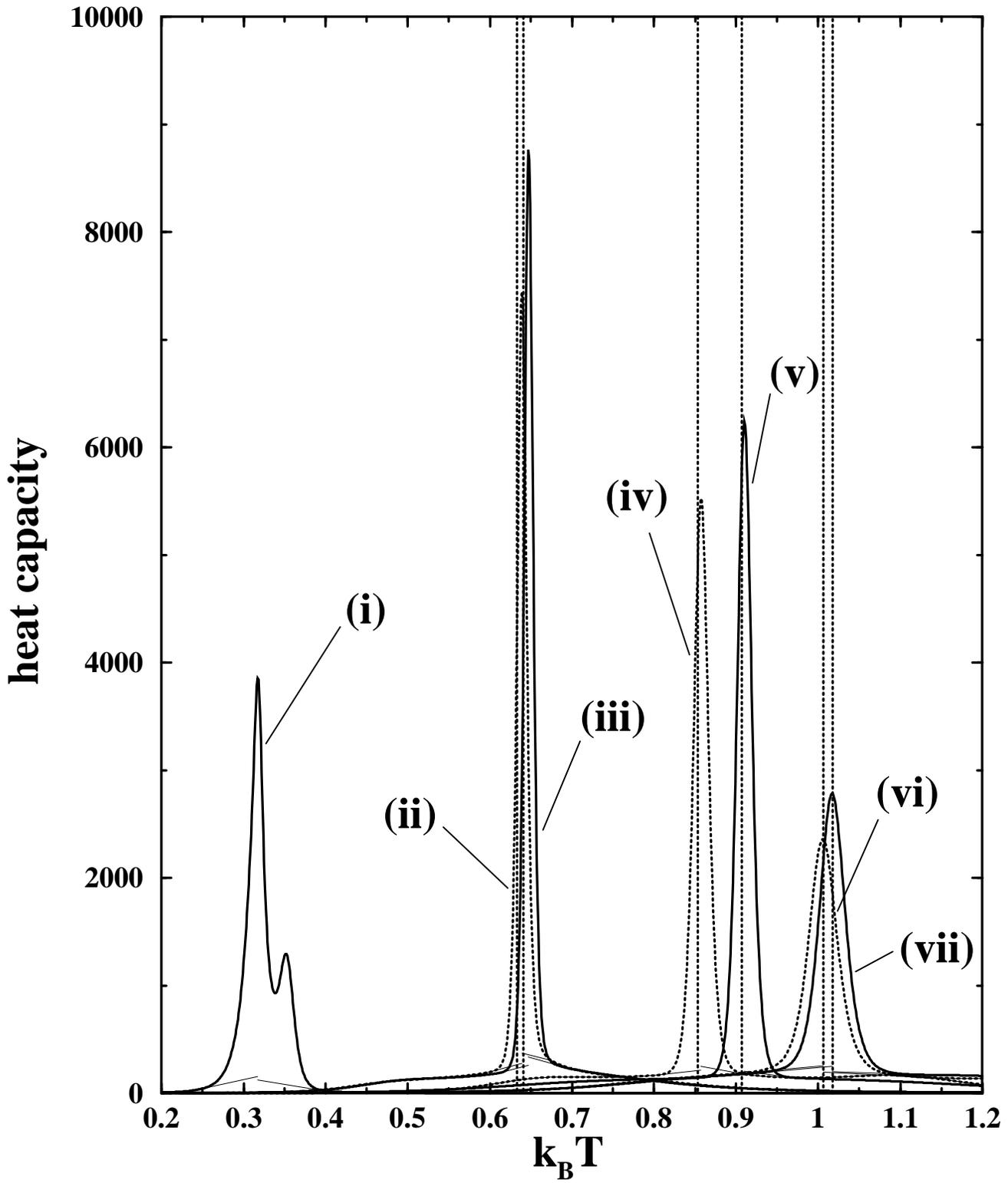
{\bf Figure 5.} $\quad$ 
Thermodynamic cooperativity.
Heat capacity as a function of temperature is shown for seven models:
(i) the contact-dominant model described in the text, (ii) NCS1 and 
(iii) NCS2 without-solvation-SSR models, (iv) NCS1 and (v) NCS2 
with-solvation models, and (vi) NCS1 and (vii) NCS2 without-solvation models.
Vertical dotted lines correspond 
to the transition midpoints marked in Figures 7--9. The computed
van't Hoff to calorimetric enthalpy ratios (defined in ref.~47)
for these models with no baseline subtractions are, respectively, 
$\kappa_2=$ $0.30$, $0.57$, $0.61$, $0.56$, $0.63$, $0.46$, and $0.50$. 
The corresponding ratios after subtracting the empirical baselines 
(shown in the figure) are $\kappa_2^{\rm (s)}=$ 
$0.33$, $0.98$, $1.00$, $1.00$, $1.01$, $0.97$, and $0.99$.  
The unit of every heat capacity scan plotted is for interaction 
strength $\epsilon=1$. Each scan was calculated from the density of states 
estimated by histogram techniques.$^{32,47}$ The sampling simulations were 
conducted at temperature $T$ and $\epsilon$ values chosen around each model's
transition midpoint to efficiently cover both the folded and unfolded 
regions of the conformational space. For (i) -- (vii), 
$T=$ $0.32$, $0.64$, $0.64$, $0.82$, $0.82$, $0.82$, and $0.82$,
respectively, and $\epsilon=$ $1.0$, $1.0$, $1.0$, $0.955$, $0.90$, 
$0.81$, and $0.81$ were used.
\\

\noindent
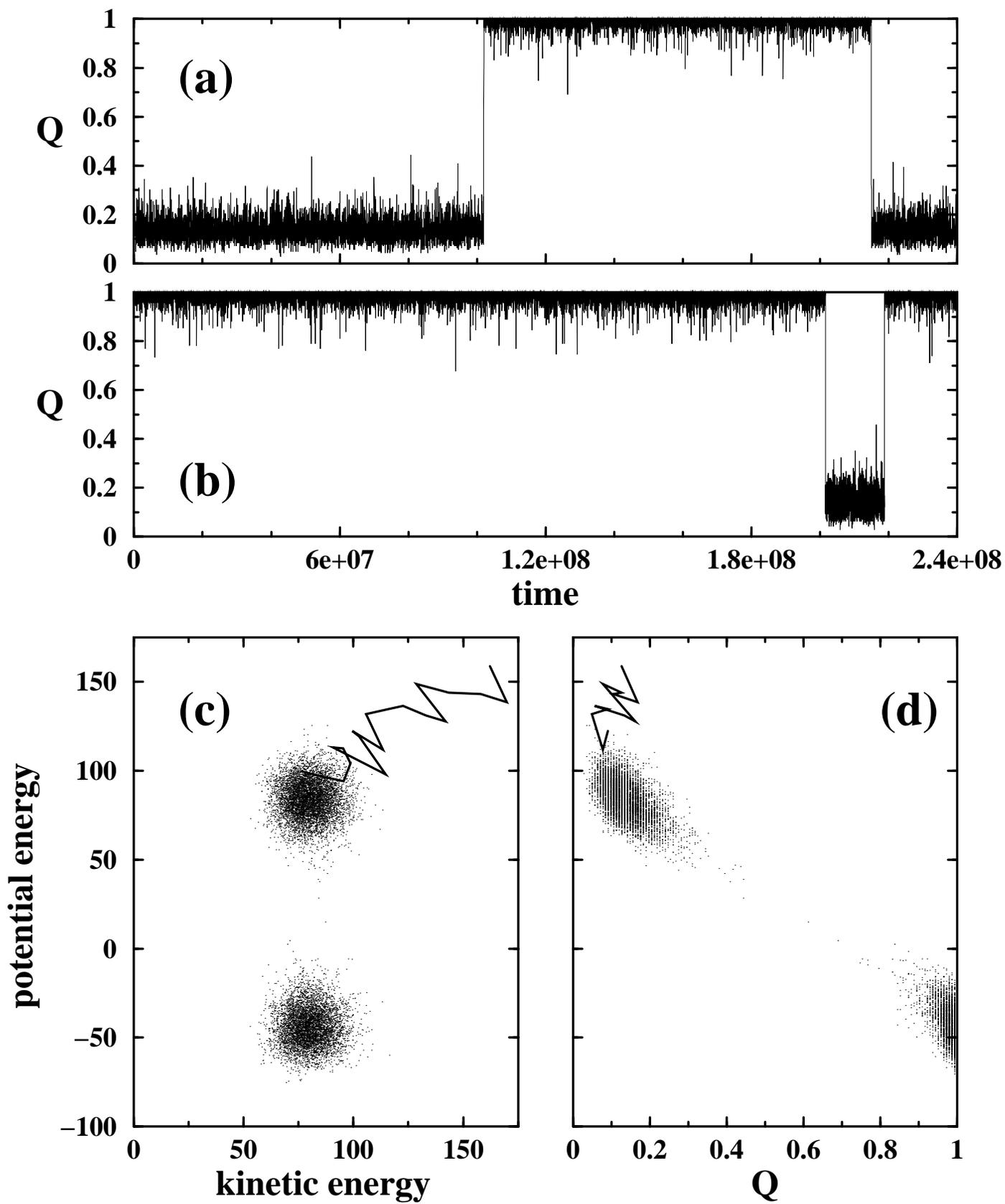
{\bf Figure 6.} $\quad$
Signatures of two-state thermodynamics and sharp kinetic 
transitions between states in the with-solvation NCS2 model
at $T=0.82$. Time evolution is monitored by snapshots taken at every 
$400\delta t$ during the simulations. (a) A folding/unfolding trajectory near 
the transition midpoint of this model at $\epsilon=0.90$ 
($-\epsilon/k_{\rm B} T=-1.098$, $\Delta G_{\rm u}/k_{\rm B} T=0.68$). 
(b) A trajectory showing transient unfolding under moderately native conditions
at $\epsilon=0.92$ ($-\epsilon/k_{\rm B} T=-1.122$, 
$\Delta G_{\rm u}/k_{\rm B} T=2.70$). 
(c) Scatter plot of potential energy $V_{\rm total}^{(S)}$ versus 
kinetic energy of the model protein (sum of $mv^2/2$ over all C$_\alpha$
positions) for the trajectory in (a). Each dot represents a snapshot. The 
first 11 snapshots are connected by line segments to highlight the initial 
pre-equilibration process. The average kinetic energy is equal to $78.9$. 
(d) is the corresponding scatter plot of potential energy versus $Q$ for 
the trajectory in (a) and (c).
\\

\noindent
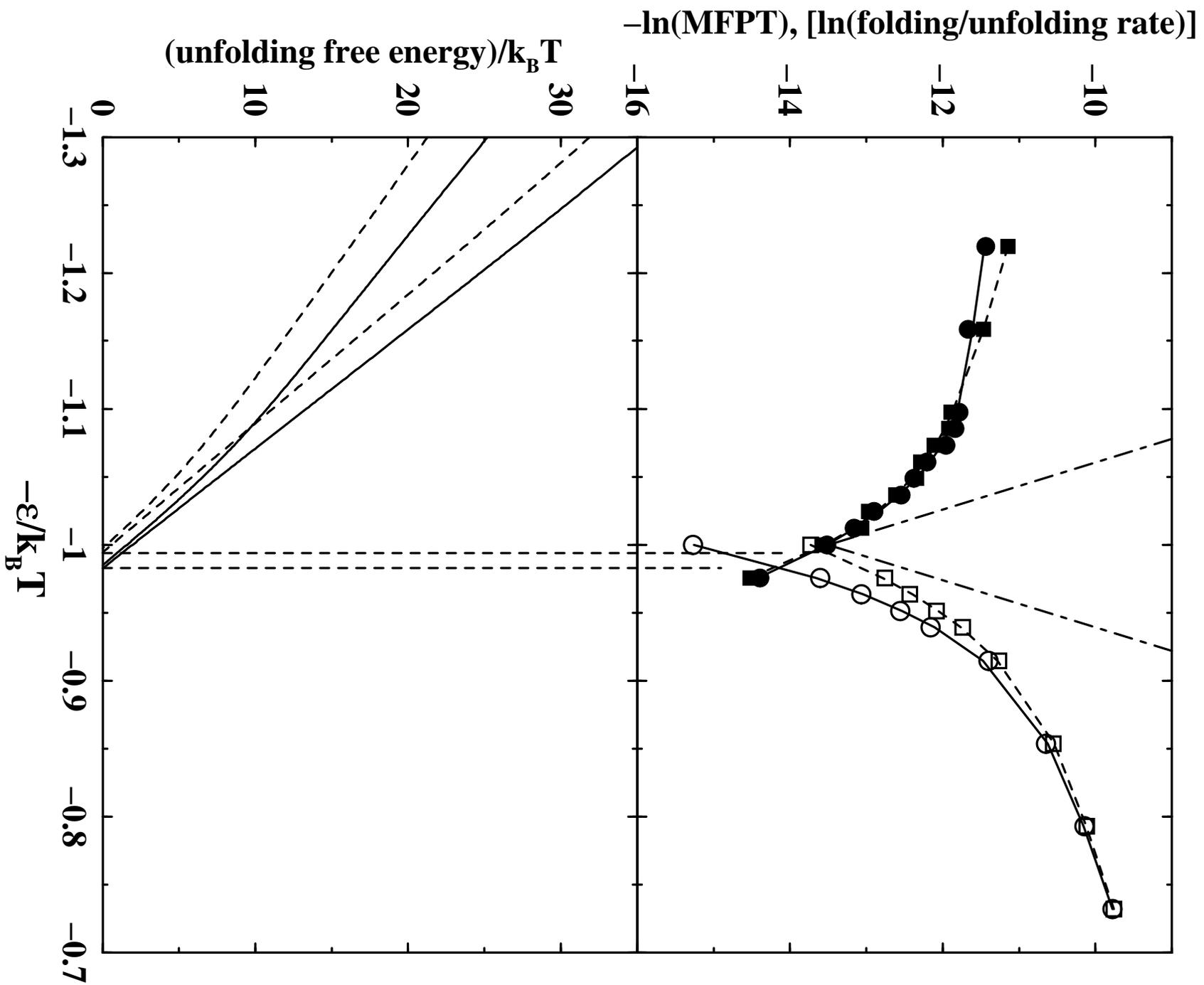
{\bf Figure 7.} $\quad$
Folding/unfolding kinetics and thermodynamics of the without-solvation 
NCS1 (squares, dashed curves) and NCS2 (circles, solid curves) 
models at $T=0.82$. {\bf Upper panel:} Chevron plots of negative natural
logarithm of folding (filled symbols) and unfolding (open symbols) 
MFPT data from Table~1. The curves are guides for the eye. Unfolding 
simulations start with the native conformation; FPT is defined by the chain 
having $\le 25$ native contacts. Folding simulations start with 
randomly generated open conformations with $Q\approx 10\%$, FPT
is defined by the chain achieving a $Q$ value larger than or equal to 
that of the native free energy minima on the free energy profiles 
in Figure~4A, i.e., $Q\ge 112/137$ for NCS1 and
$Q\ge 120/142$ for NCS2. 
{\bf Lower panel:} The free energy of unfolding $\Delta G_{\rm u}$ 
in units of $k_{\rm B} T$ for each of the two models (dashed lines: NCS1,
solid lines: NCS2)
is the natural logarithm of the Boltzmann weight (population) of the folded
state minus that of the denatured chain population with $\le 35$ native 
contacts (upper curves) or that with $\le 70$ native contacts (lower curves).
Conformations with $\ge 100$ out of 137, and $\ge 105$ out of 142 
native contacts (corresponding approximately to $Q\ge 0.73$ around
the native minima in Figure~4A) are taken 
to be the folded states, respectively, of the NCS1 and NCS2 models. 
Stability curves shown are obtained by histogram techniques from 
simulations at $\epsilon=0.80$ and $0.81$ for NCS1, and at $\epsilon=0.80$, 
$0.81$, and $0.82$ for NCS2. The vertical dotted line marks the midpoint
$-\epsilon/k_{\rm B} T$ values at which $\Delta G_{\rm u}=0$ for the two 
models.  The V-shaped dashed-dot lines in the upper panel is
an hypothetical simple two-state chevron plot that would be consistent 
with the models' approximately linear thermodynamic stability curves 
in the lower panel. Note that the qusai-linear stability curves of
the two models have approximately the same slope. 
\\

\noindent
\noindent
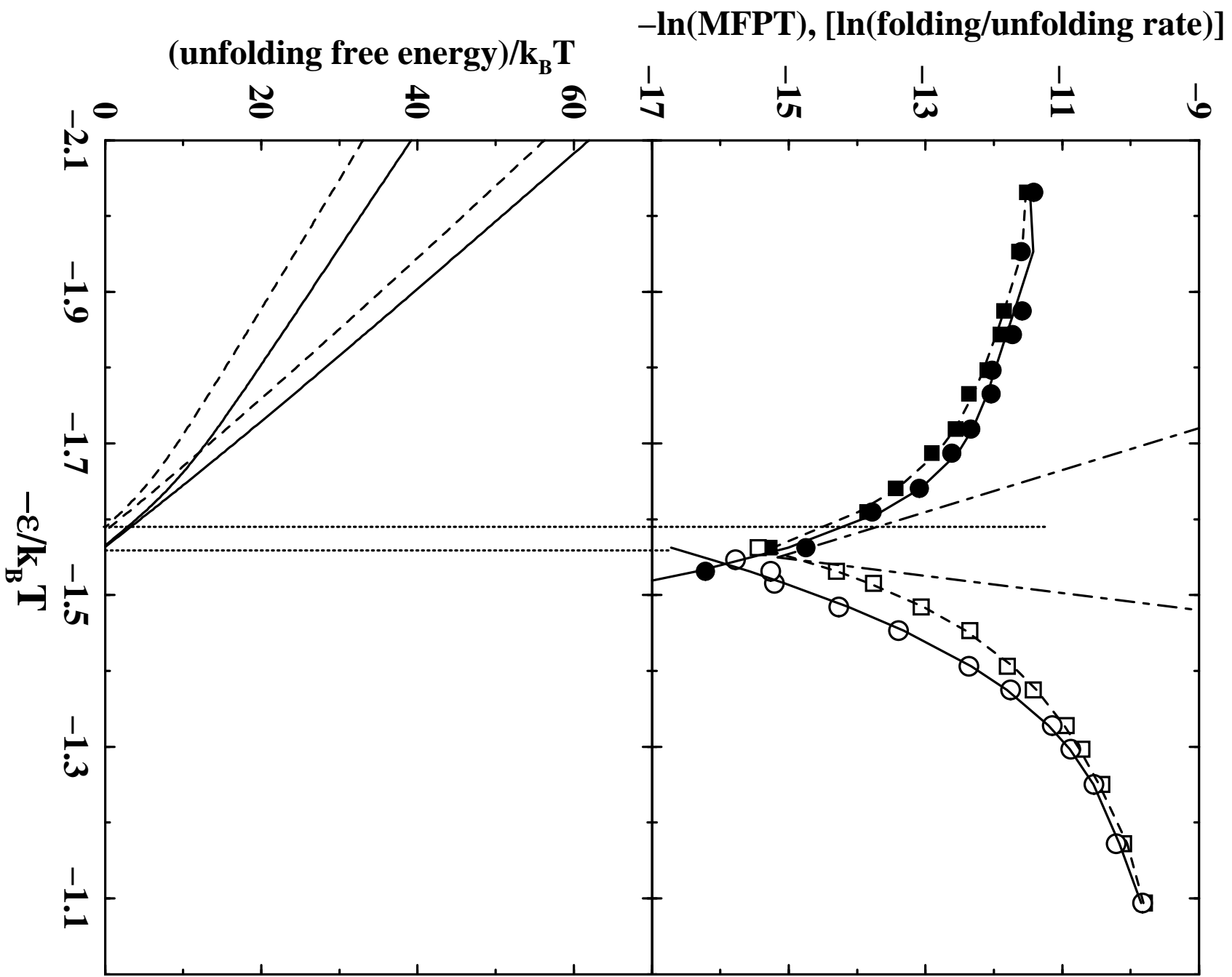
{\bf Figure 8.} $\quad$
Same as Figure~7 but for the without-solvation-SSR models at $T=0.64$. 
Simulation details not identical to that in Figure~7 are 
as follows. {\bf Upper panel:} MFPTs are from Table~2.
Here folding FPT is defined by the chain achieving $Q=1$.
{\bf Lower panel:} Stability curves 
are given by the natural logarithm of the Boltzmann weight (population) 
of the folded state minus that of the denatured chain population with 
$\le 35$ native contacts (upper curves) or that with $\le 80$ native 
contacts (lower curves).  Conformations with 132 out of 137, and 137 
out of 142 native contacts (corresponding to the native minima on the 
free energy profiles in Figure~4B) are taken to be the folded states, 
respectively, of the NCS1 and NCS2 models. The $\Delta G_{\rm u}/k_{\rm B} T$ 
stability curves remain essentially unchanged if the thermodynamic definitions
for the folded states of these models are extended to $Q\ge 132/137$ 
for NCS1 and $Q\ge 137/142$ for NCS2. Stability curves shown are 
obtained by histogram techniques from simulations at $\epsilon=0.97$,
$0.98$, $0.99$, and $1.00$ for NCS1, and at $\epsilon=0.99$, $1.00$, 
and $1.01$ as well as confirmed by simulations at several temperatures 
other than $T=0.64$ for NCS2. 
\\

\noindent
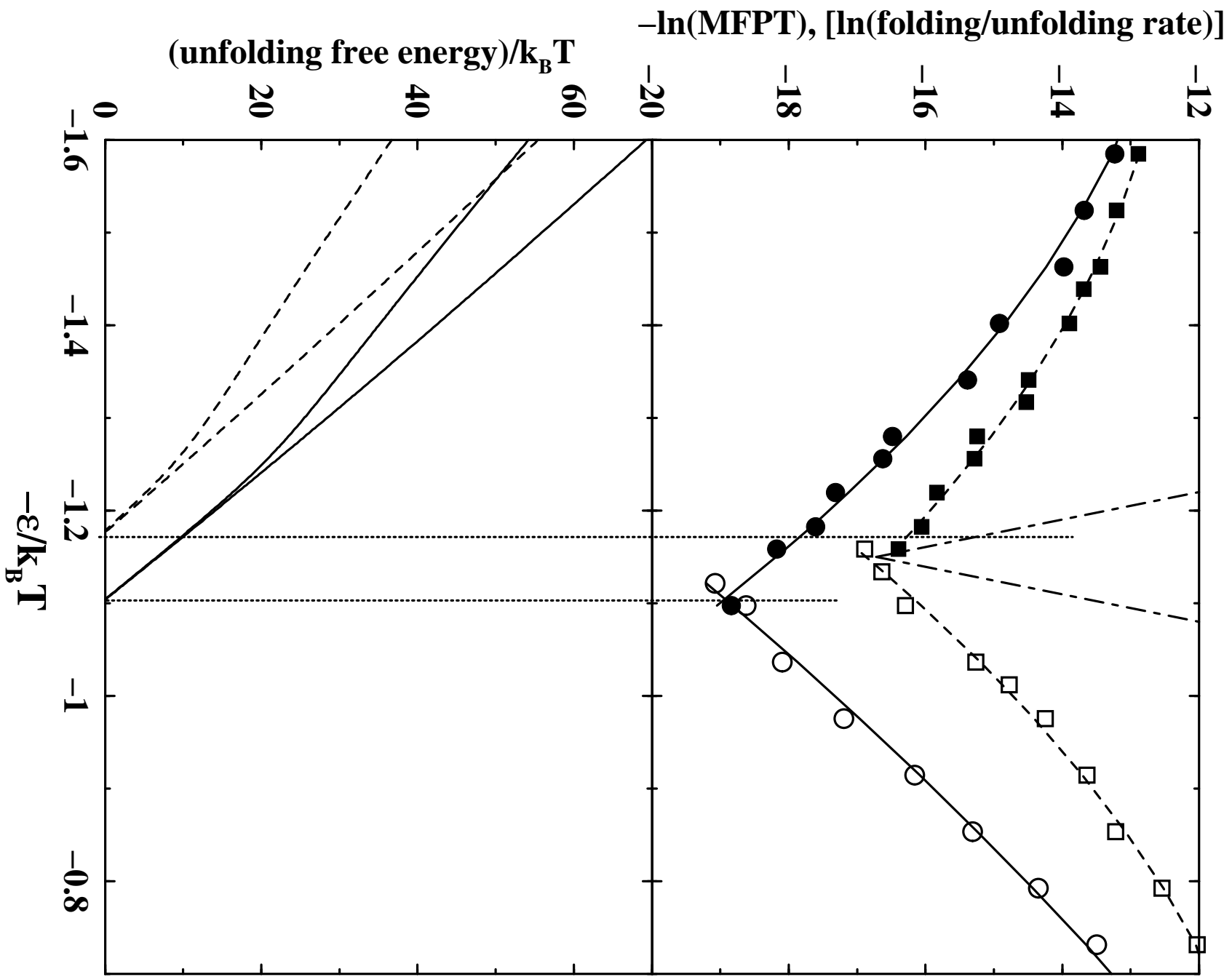
{\bf Figure 9.} $\quad$
Same as Figure~8 but for the with-solvation models at $T=0.82$.
Simulation details not identical to that in Figure~8 are 
as follows. {\bf Upper panel:} MFPTs are from the 
$0.6\le\epsilon\le 1.30$ entries in Table~3. Otherwise the kinetic 
definitions of folding and unfolding are the same as that in Figure~8.
{\bf Lower panel:} Stability curves 
are given by the natural logarithm of the Boltzmann weight of the folded state
minus that of the denatured chain population with $\le 30$ native 
contacts (upper curves) or that with $\le 80$ (lower curves).
The folded state is defined here by conformations with exactly $Q=1$ 
(c.f. Figure~4C). The stability curves are obtained
by histogram techniques from simulations at $\epsilon=0.955$ and 
$0.96$ for NCS1, and at $\epsilon=0.90$ for NCS2. 
\\

\noindent
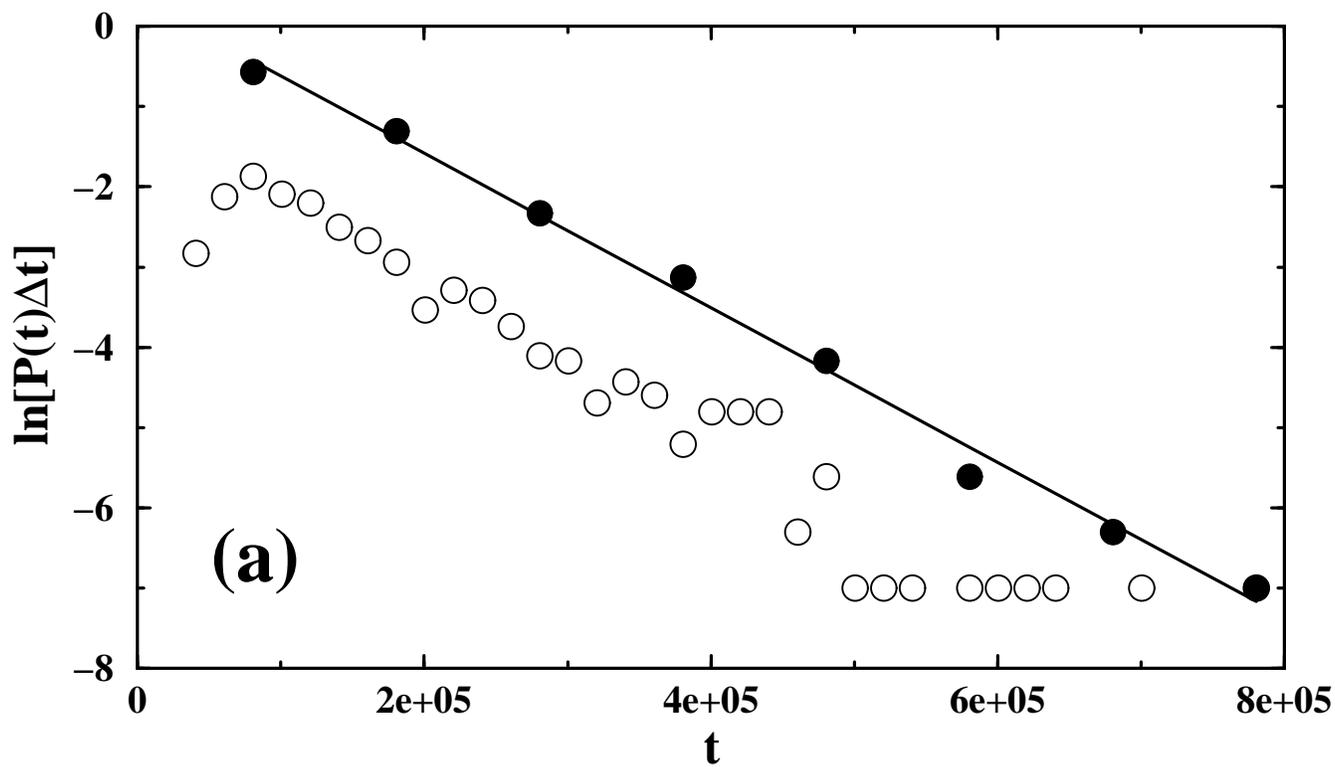
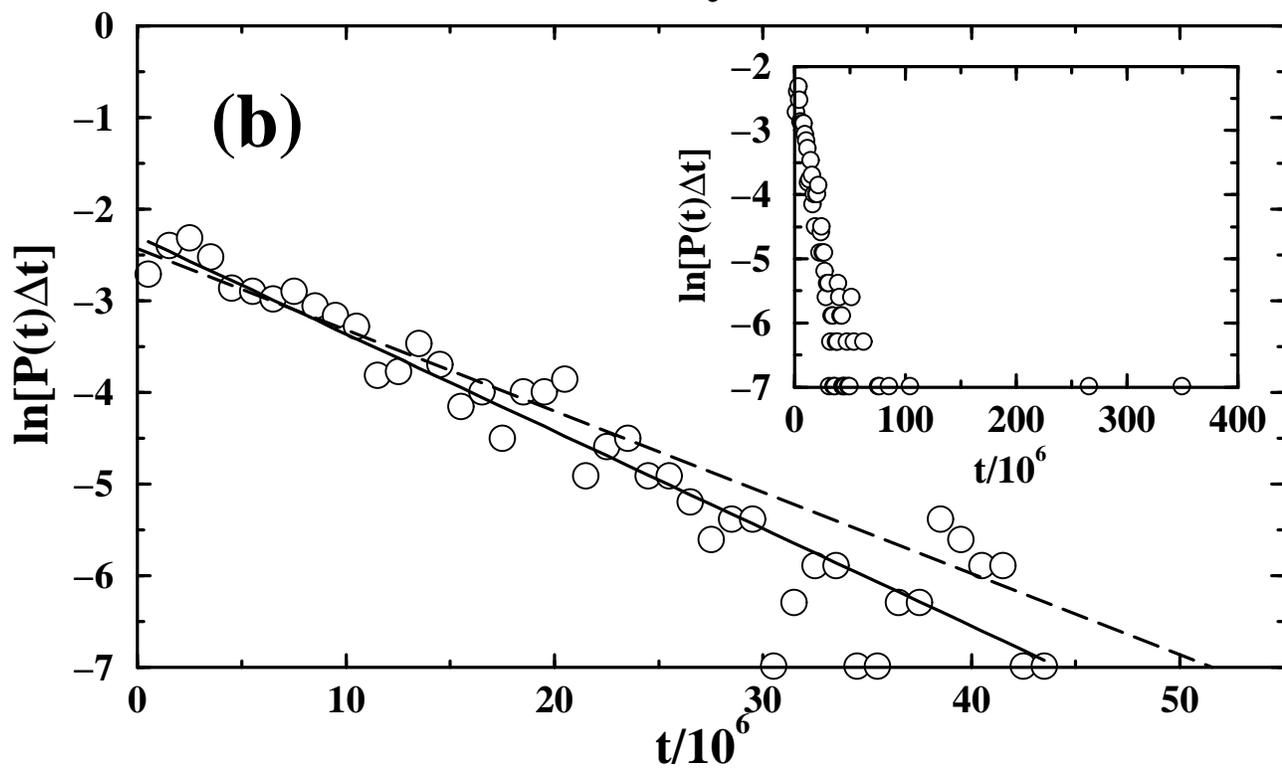
{\bf Figure 10.} $\quad$
Approximate single exponential folding kinetics indicated by first 
passage time (FPT) distributions. $P(t)\Delta t$ is the fraction of 
trajectories with $t-\Delta t/2<$ FPT $\le t+\Delta t/2$.
(a) The FPT distribution among the 1,097 folding trajectories
under strongly native conditions at $-\epsilon/k_{\rm B} T=-1.80$ in the 
NCS2 without-solvation-SSR model ($\epsilon=1.15$ entry in Table~2)
is shown for bin sizes $\Delta t= 10^5$ (filled circles) and
$\Delta t= 2\times 10^4$ (open circles). 
The solid line is the least-square fit through the $\Delta t= 10^5$ 
data points. (b) Included for comparison is the FPT distribution
among the 1,080 folding trajectories of the three-dimensional lattice 
model described on page~903 of Kaya and Chan$^{68}$ with 
$\epsilon/k_{\rm B} T =-1.72$ and $\Delta t= 10^6$; 
$t$ is the number of Monte Carlo time steps.
The solid line is the least-square fit (correlation coefficient $r=0.95$)
through the data points shown in the main figure; 20 trajectories that 
give rise to a long-FPT tail in the full distribution (inset) are excluded 
from this fit. The dashed line is equation~6 with $\ln({\rm MFPT})=16.25$ 
from ref.~68; $t_0$ was taken to be zero for this lattice case.
\\

\noindent
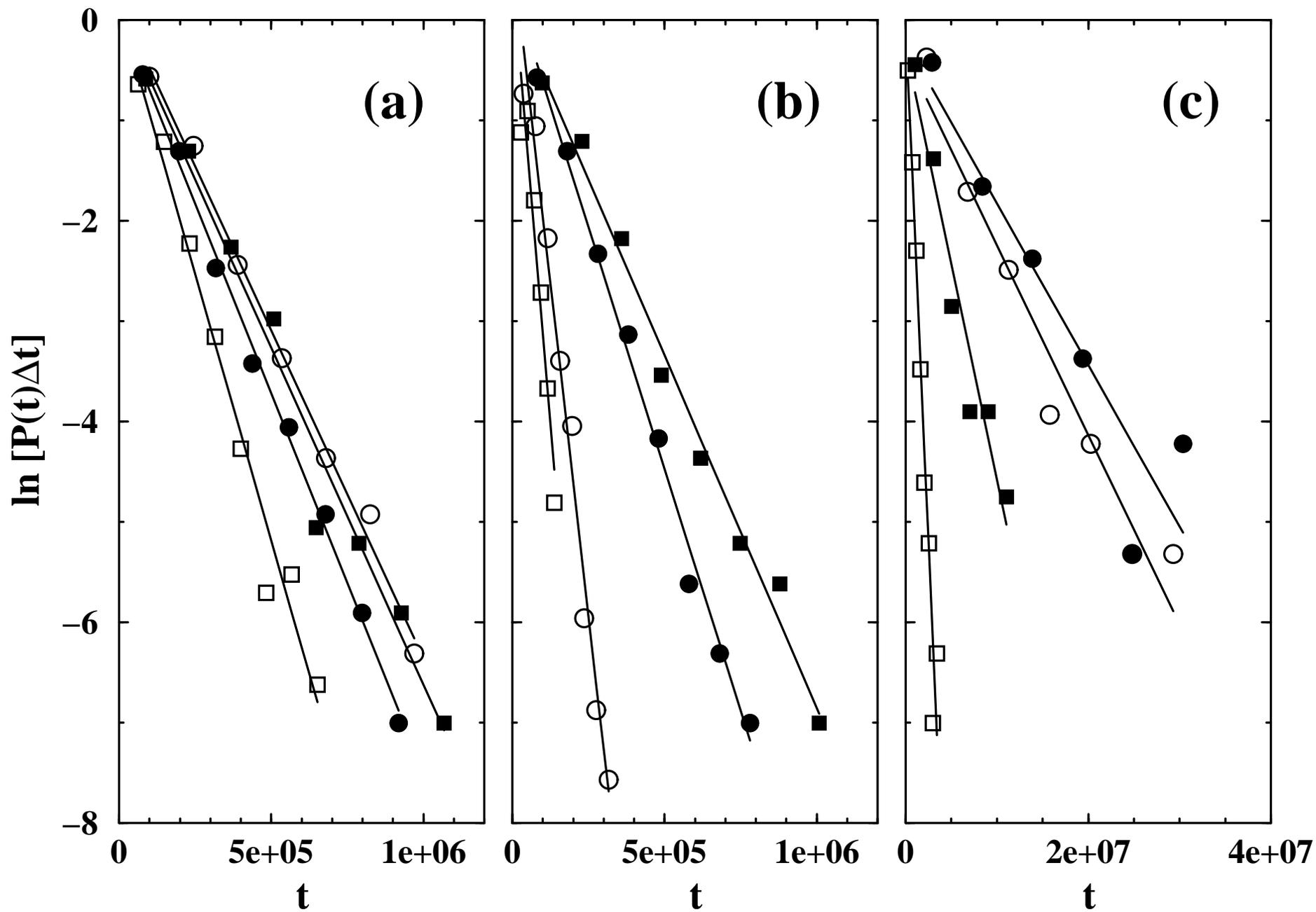
{\bf Figure 11.} $\quad$
Approximate single-exponential folding and unfolding kinetics.
FPT distributions are presented as in Figure~10A. Solid lines are 
least-square fits through the data points shown. Numbers of trajectories 
in the distributions are given in Tables~1--3. Unfolding and folding 
data using NCS1 (or NCS2) are plotted, respectively, by open and filled 
squares (or circles). (a) Without-solvation models. 
(b) Without-solvation-SSR models. The NCS2 folding plot here is identical 
to the $\Delta t=10^5$ case in Figure~10A. (c) With-solvation models.
The $\epsilon$ values for NCS1 unfolding, folding, NCS2 
unfolding, folding, and the corresponding $\Delta t$ bin sizes for
these different models are, respectively, 
(a) $\epsilon=$ $0.77$, $0.88$, $0.77$, $0.88$, 
$\Delta t/10^5=$ $0.84$, $1.4$, $1.5$, $1.2$; (b) $\epsilon=$ $0.85$, 
$1.15$, $0.85$, $1.15$, $\Delta t/10^5=$ $0.21$, $1.3$, $0.4$, $1.0$; 
(c) $\epsilon=$ $0.70$, $1.10$, $0.70$, $1.10$, 
$\Delta t/10^6=$ $0.45$, $2.0$, $4.6$, $5.5$.
\\

\noindent
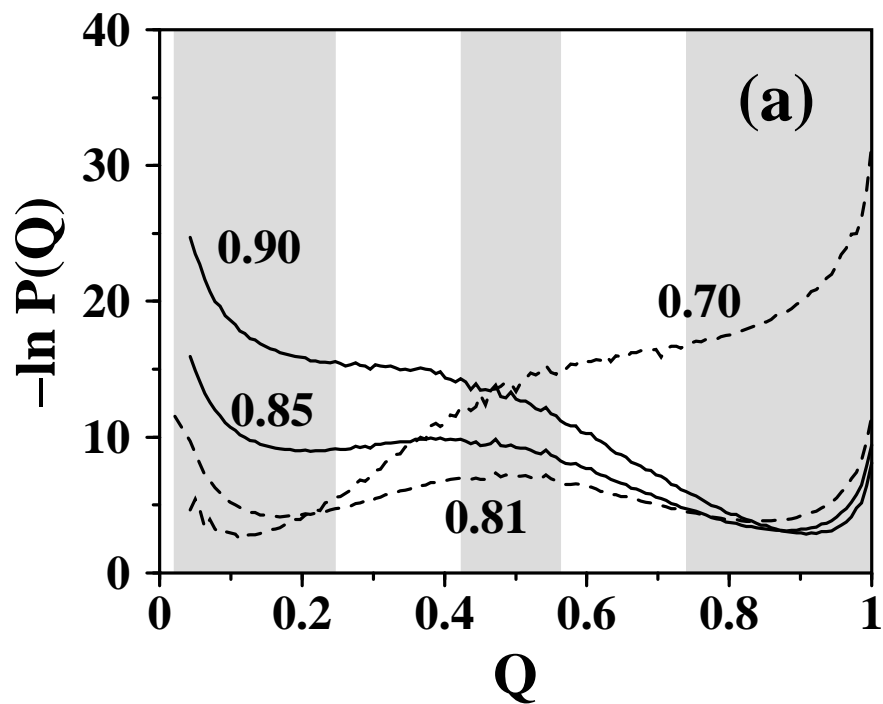
{\bf Figure 12.} $\quad$
Front factor analyses.
(a -- c): Free energy profiles for the NCS2 without-solvation 
(a), without-solvation-SSR (b) and with-solvation (c) models at 
the $\epsilon$ values indicated (c.f. Figure~4). 
These plots are obtained from Boltzmann distributions 
computed by histogram techniques. Where
appropriate, different line styles are used for profiles at different 
$\epsilon$ values for clarity. Shaded areas are examples of unfolded,
transition, and folded state regions considered (see text).
(d) Correlations between rates and activation free energies
deduced from free energy profiles are analyzed as in ref.~68. The vertical 
variable is given by $\ln F$ $=$ $-\ln ({\rm MFPT})$ 
$+$ $\Delta G^\ddagger/k_{\rm B} T$ (equation~7). In the present 
analysis, $\Delta G^\ddagger(\epsilon,T)/k_{\rm B} T=$
$-\ln [P(60/142\le Q \le 80/142)/P(Q\ge 105/142)]$
(open triangles, without-solvation unfolding),
$-\ln [P(60/142\le Q \le 80/142)/P(Q\le 35/142)]$
(filled triangles, without-solvation folding),
$-\ln [P(75/142\le Q \le 95/142)/P(Q=137/142)]$
(open squares, without-solvation-SSR unfolding),
$-\ln [P(75/142\le Q \le 95/142)/P(Q\le 25/142)]$
(filled squares, without-solvation-SSR folding),
$-\ln [P(70/142\le Q \le 95/142)/P(Q=1)]$
(open circles, with-solvation unfolding), and
$-\ln [P(70/142\le Q \le 95/142)/P(Q\le 25/142)]$
(filled circles, with-solvation folding).
The middle shaded regions in (a -- c) correspond to the 
transition-state regions used in the analysis in (d). NCS1 models 
have similar trends (data not shown).         
\\

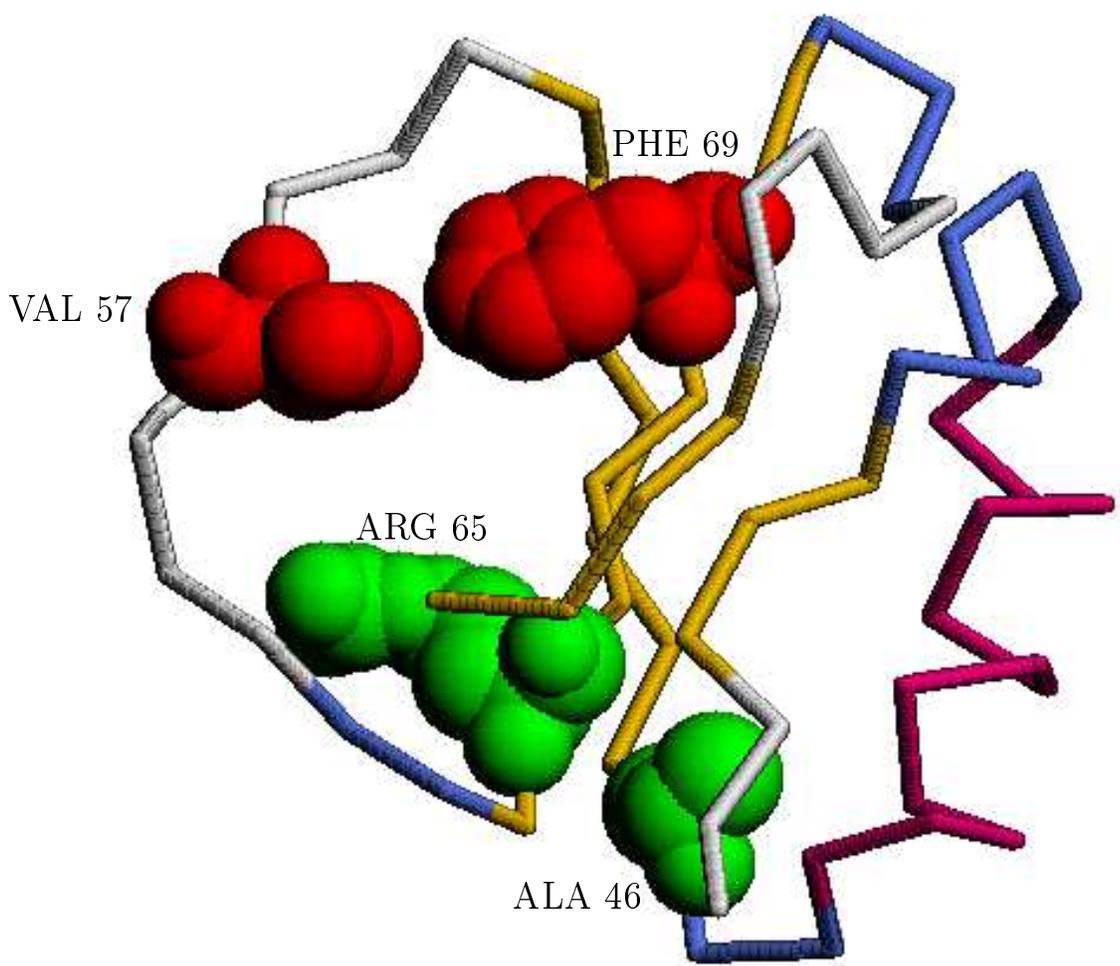
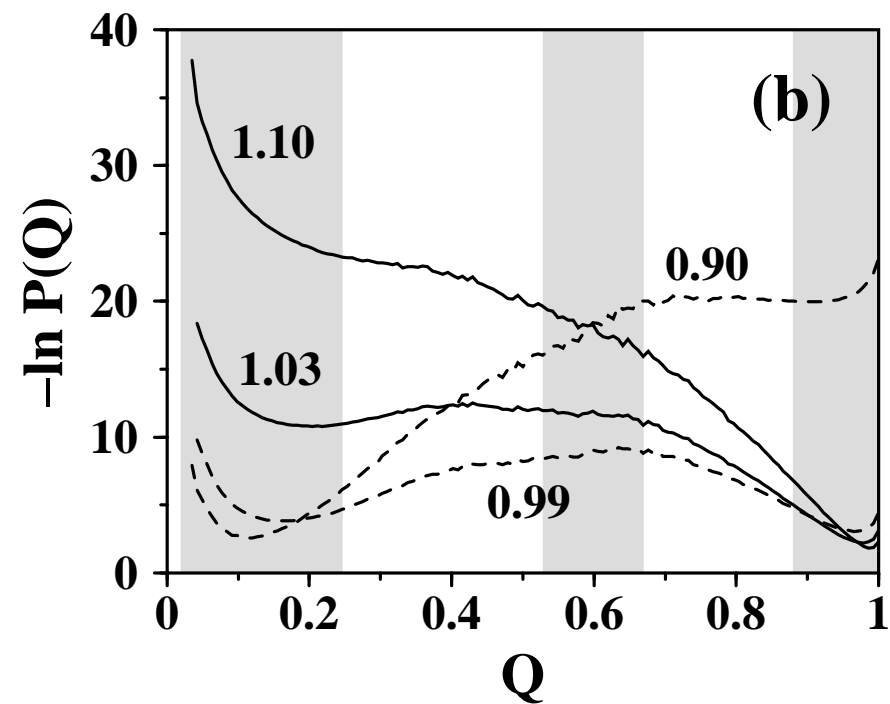
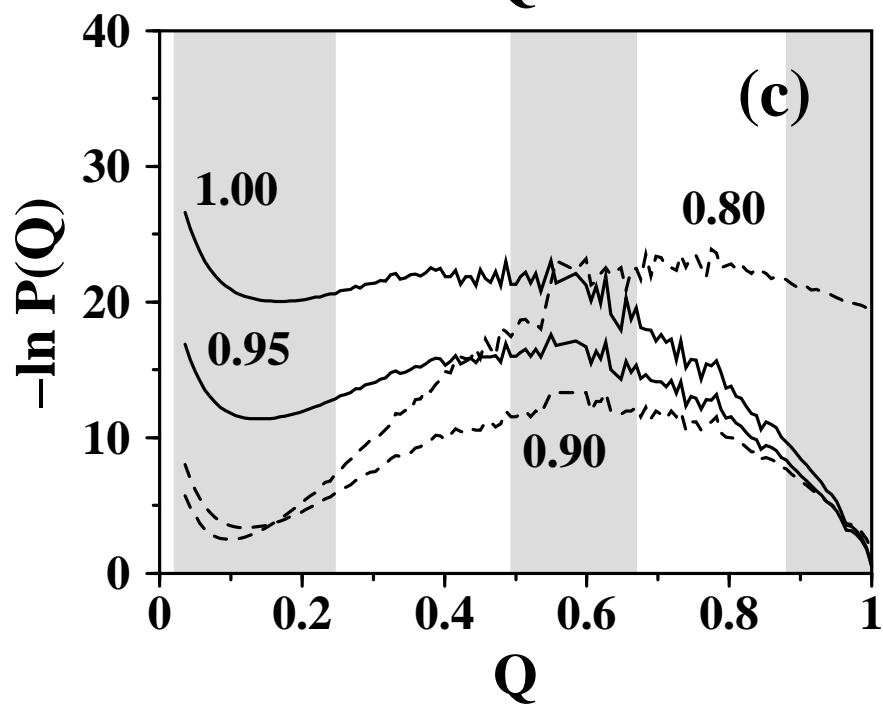
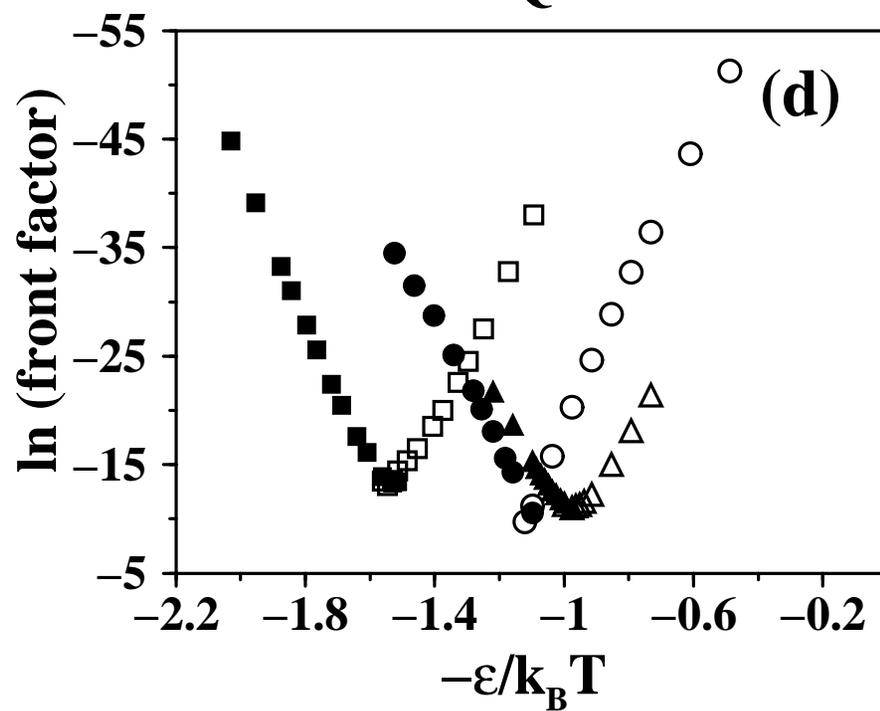
\end{document}